\documentclass{aastex61}

\accepted{May 3, 2017}
\submitjournal{ApJ}

\shortauthors{Neeley et al.}

\begin{document}

\title{On a new theoretical framework for RR Lyrae stars II: 
Mid--Infrared Period--Luminosity--Metallicity Relations}

\correspondingauthor{Jillian R. Neeley}
\email{jrneeley@iastate.edu}

\author{Jillian R. Neeley}
\affil{Department of Physics \& Astronomy, Iowa State University, Ames, IA 50011, USA}

\author{Massimo Marengo}
\affiliation{Department of Physics \& Astronomy, Iowa State University, Ames, IA 50011, USA}

\author{Giuseppe Bono}
\affiliation{Department of Physics, Universit\`a di Roma Tor Vergara, via della Ricerca Scientifica 1, 00133 Roma, Italy}
\affiliation{INAF-Osservatorio Astronomico di Roma, via Frascati 33, I-00040 Monte Porzio Catone, Italy}

\author{Vittorio F. Braga}
\affiliation{Department of Physics, Universit\`a di Roma Tor Vergara, via della Ricerca Scientifica 1, 00133 Roma, Italy}
\affiliation{ASDC, via del Politecnico snc, 00133 Roma, Italy}
\affiliation{INAF-Osservatorio Astronomico di Roma, via Frascati 33, I-00040 Monte Porzio Catone, Italy}
\affiliation{Departamento de Fisica, Universidad Andres Bello, Av. Fernandez Concha 700, Las Condes, Santiago, Chile}
\affiliation{Instituto Milenio de Astrofisica, Santiago, Chile}

\author{Massimo Dall'Ora}
\affiliation{INAF-Osservatorio Astronomico di Capodimonte, Salita Moiarello 16, I-80131 Napoli, Italy}

\author{Davide Magurno}
\affiliation{Department of Physics, Universit\`a di Roma Tor Vergara, via della Ricerca Scientifica 1, 00133 Roma, Italy}
\affiliation{INAF-Osservatorio Astronomico di Roma, via Frascati 33, I-00040 Monte Porzio Catone, Italy}

\author{Marcella Marconi}
\affiliation{INAF-Osservatorio Astronomico di Capodimonte, Salita Moiarello 16, I-80131 Napoli, Italy}

\author{Nicolas Trueba}
\affiliation{Department of Physics \& Astronomy, Iowa State University, Ames, IA 50011, USA}

\author{Emanuele Tognelli}
\affiliation{Dipartimento di Fisica, Università di Pisa, Lago Bruno Pontecorvo 3, I-56127, Pisa, Italy}
\affiliation{INFN, Sezione di Pisa, Lago Bruno Pontecorvo 3, I-56127, Pisa, Italy}

\author{Pier G. Prada Moroni}
\affiliation{Dipartimento di Fisica, Università di Pisa, Lago Bruno Pontecorvo 3, I-56127, Pisa, Italy}
\affiliation{INFN, Sezione di Pisa, Lago Bruno Pontecorvo 3, I-56127, Pisa, Italy}

\author{Rachael L. Beaton}
\affiliation{Carnegie Observatories, 813 Santa Barbara Street, Pasadena, CA 91101, USA}

\author{Wendy L. Freedman}
\affiliation{Department of Astronomy \& Astrophysics, University of Chicago, Chicago, IL 60637, USA}

\author{Barry F. Madore}
\affiliation{Carnegie Observatories, 813 Santa Barbara Street, Pasadena, CA 91101, USA}

\author{Andrew J. Monson}
\affiliation{Department of Astronomy \& Astrophysics, The Pennsylvania State University, 525 Davey Lab, University Park, PA 16802, USA}

\author{Victoria Scowcroft}
\affiliation{Department of Physics, University of Bath, Claverton Down, Bath, BA2 7AY, United Kingdom}

\author{Mark Seibert}
\affiliation{Carnegie Observatories, 813 Santa Barbara Street, Pasadena, CA 91101, USA}

\author{Peter B. Stetson}
\affiliation{NRC-Herzberg, Dominion Astrophysical Observatory, 5071 West Saanich Road, Victoria BC V9E 2E7, Canada}



\begin{abstract}

We present new theoretical period-luminosity-metallicity (PLZ) relations for RR Lyr\ae{} stars (RRL) at \emph{Spitzer} and \emph{WISE} wavelengths. The PLZ relations were derived using nonlinear, time-dependent convective hydrodynamical models for a broad range in metal abundances (Z=0.0001 to 0.0198). In deriving the light curves, we tested two sets of atmospheric models \citep{BrottHauschildt2005, CastelliKurucz2003} and found no significant difference between the resulting mean magnitudes. We also compare our theoretical relations to empirical relations derived from RRL in both the field and in the globular cluster M4. Our theoretical PLZ relations were combined with multi-wavelength observations to simultaneously fit the distance modulus, $\mu_0$, and extinction, $A_V$, of both the individual Galactic RRL and of the cluster M4. The results for the Galactic RRL are consistent with trigonometric parallax measurements from \emph{Gaia's} first data release. For M4, we find a distance modulus of $\mu_0=11.257\pm 0.035$ mag with $A_V = 1.45\pm 0.12$ mag, which is consistent with measurements from other distance  indicators. This analysis has shown that when considering a sample covering a range of iron abundances, the metallicity spread introduces a dispersion in the PL relation on the order of 0.13 mag. However, if this metallicity component is accounted for in a PLZ relation, the dispersion is reduced to $\sim 0.02$ mag at MIR wavelengths. 

\end{abstract}




\section{Introduction} \label{sec:intro}

RR Lyr\ae{} (RRL) variables are a popular tracer for old stellar populations, thanks to their abundance in the globular clusters, halos and bulges of galaxies (see e.g. \citealt{vivas2006, dekany2013, pietrukowicz2015}). During their advanced evolutionary phases low and intermediate-mass stars cross the so-called Cepheid instability strip (a region of the HR diagram in which stellar atmospheres are pulsationally stable). Unlike their higher mass counterparts, RRLs do not appear to obey well defined period-luminosity (PL) relations at visible wavelengths. As a consequence, their usefulness as distance indicators has been historically limited to the adoption of a luminosity-metallicity relation characterized by a rather large ($\approx 5$\%) intrinsic scatter \citep{caceres2008}. This relation also suffers from evolutionary effects and uncertainties related to the metallicity scale and/or $\alpha$ enhancement. Moreover, the relation is possibly nonlinear across the whole observed RRL metallicity range \citep{caputo2000}. 

Several theoretical and empirical arguments, however, indicate that RRLs become solid distance indicators when moving from the optical to the infrared bands. As described as length in \citet{bono2016b}, the main reasons are three-fold: 

\renewcommand{\theenumi}{\roman{enumi}}
\begin{enumerate}

\item As first demonstrated over 30 years ago by \citet{longmore1986} and again by \citet{dallora2004}, an obvious PL relation does appear moving from the optical to the infrared bands. While the slope is vanishingly small in the $V$-band, it becomes steeper with increasing wavelength \citep{catelan2004, marconi2015}, ranging from $-1.2$ in the $R$-band to $-2.2$ in the $K$-band. This behavior is different than the one shown by Cepheids \citep{bono1999} and is related to the specific dependence of the RRL bolometric correction with temperature \citep{bono2001, bono2003, bono2003aspc}. For $\lambda \ga 2.2 \ \mu$m the slope of the PL relations of instability strip pulsators (both Cepheids and RRLs) becomes almost constant. In this wavelength regime the brightness variations of pulsating stars is mainly driven by their radius variation, and the effective temperature variations only play a minimal role \citep{jameson1986, madore2012}.

\item The intrinsic dispersion of the RRL PL relations steadily decreases when moving from the optical to the infrared bands. This trend is due to the fact that starting from the near-infrared (NIR) cooler RRLs  are steadily brighter than hotter ones, due to the stronger temperature sensitivity of the bolometric correction \citep{bono2003aspc}. As a consequence, infrared PL relations are only marginally affected by the intrinsic width in  temperature of the instability strip, since they almost mimic a period-luminosity-color relation. Furthermore, the instability strip itself becomes narrower at longer wavelengths, and as a consequence the color term responsible for the intrinsic dispersion on the PL relations vanishes \citep{catelan2004, madore2012, marconi2015}. This means that at MIR wavelengths PL relations provide for more precise distance indicators \citep{braga2015}.

\item Infrared observations are less affected by reddening, due to the power-law dependency of the extinction laws ($\lambda^{-\beta}$, with $\beta \sim 1.6$ to $1.8$; \citealt{bono2016b}). This translates into a smaller uncertainty of reddening and differential reddening, when compared to the $V$-band, by a factor ranging from four ($J$) to ten ($K$).  In the $L$ and $M$ bands extinction is further reduced, reaching its minimum values with $A_V/A_\lambda \approx 15$ and 20, respectively. This is a huge advantage when inspecting highly- and differentially-reddened targets, such as those in the Galactic bulge \citep{nishiyama2009}.

\end{enumerate}

To explore the detailed physics behind these remarkable properties, \citet{marconi2015} used new, time- and metal-dependent convective hydrodynamic models to derive a theoretical calibration for the period-luminosity-metallicity (PLZ), period-Wesenheit-metallicity (PWZ) and metal-independent period-Wesenheit (PW) relations of RRL. These relations have been published in the optical (Johnson-Kron-Cousins's $BVRI$) and NIR (2MASS $JHK$) wavelength regimes \citep{marconi2015}, and tested by fitting average magnitudes of RRLs in the M4 (NGC 6121) Galactic globular cluster \citep{braga2015} and in the dwarf spheroidal galaxy Carina \citep{coppola2015}. On the other hand, similar analysis in the mid-infrared (MIR), where the extinction is lower and the intrinsic temperature-dependent scatter is at its smallest, is lagging. Several authors \citep{klein2011, madore2013, neeley2015} have derived empirical calibrations for Galactic RRL MIR PL relations, but their zero point calibration is based on just five stars for which $\approx 10$\% accuracy Hubble Space Telescope (HST)  Fine Guidance Sensor (FGS) parallaxes are available \citep{benedict2011}. A detailed theoretical analysis of RRL PLZ relations in the MIR is missing. To fill this gap we decided to extend the detailed investigation of RRL pulsation properties provided by \citet{marconi2015} in order to derive a theoretical calibration of RRL PLZ and PWZ relations in the MIR. In this paper we focus on the MIR bands available to the InfraRed Array Camera (IRAC, \citealt{fazio2004}) onboard the Spitzer Space Telescope \citep{werner2004}, as well as the passbands of the Wide-field Infrared Survey Explorer (\emph{WISE}, \citealt{wright2010}).

Our choice of filters is motivated by two reasons. Firstly, both IRAC and \emph{WISE} provide a large archive of Galactic and extragalactic observations with thousands of RRL observed in one or multiple epochs. These include the ``warm'' Spitzer observational campaigns part of the Carnegie RR Lyrae{} Program \citep[CRRP;][]{2012sptz.prop90002F} and the Spitzer Merger History and Shape of the Halo program \citep[SMHASH;][]{2013sptz.prop10015J}, designed to obtain multi-epoch light-curves of RRLs in the halo and bulge of the Milky Way, as well as in dwarf galaxies and tidal streams. Secondly, the wavelength range covered by IRAC and \emph{WISE} will be available as part of the James Webb Space Telescope (JWST) Near Infrared Camera (NIRCam) and the Mid-Infrared Instrument (MIRI) imagers. The NIRCam wide filters F356W and F444W, in particular, are analogous to the IRAC warm passbands at 3.6 and 4.5~$\micron$. The predicted sensitivity of these two filters (13.8 and 24.5~nJy for F356W and F444W for a 10$\sigma$ detection in 10,000~sec integration\footnote{\url{http://www.stsci.edu/jwst/instruments/nircam/sensitivity/table}}) will allow the detection of RRLs at a distance modulus as high as $\sim 26$ mag (1.6 Mpc), thereby covering a significant portion of the Local Group, at least in galaxies where the $\sim 0.2$~arcsec resolution will be sufficient to beat confusion. This will open the possibility of using RRL distances to precisely calibrate secondary distance indicators independently from Cepheids, with the goal to provide a Population II route to the cosmological distance scale \citep{beaton2016}.

A detailed description of the models, and the procedure we followed to derive the PLZ and PWZ relations in the IRAC and \emph{WISE} bands, is presented in Section~\ref{chapt_intro}. To test the reliability of our theoretical relations in providing accurate distance estimates of individual stars, we have collected average magnitudes for a sample of Galactic RRLs with a broad range of metallicity, as well as revised our previously published photometry for the RRLs in the globular cluster M4. These observations are presented in Section~\ref{obs}. We compare our theoretical PLZ relations to empirical PL and PLZ relations from the literature in Section~\ref{empiricalPL}. In Sections~\ref{galactic} and \ref{M4} we use our synthetic PLZ relations to fit new distance moduli and visual band extinctions for Galactic and M4 RRL, respectively, and compare these results to other estimates. The conclusions of our work, with a focus on the dependence of the PLZ relations on metallicity, are presented in Section~\ref{concl}.


\section{Theoretical framework} \label{chapt_intro}

Theoretical models allow us to derive new theoretical MIR PLZ 
relations, by adopting the same models and following the same steps 
described in \citet{marconi2015}. For a detailed discussion on the models, 
see Section~2 of the quoted paper. Here, we want to stress only a few major 
points.
\renewcommand{\theenumi}{\roman{enumi}}
\begin{enumerate}
\item The models span a large range in metallicity. Seven
different values of $Z$ ranging from 0.0001 to 0.0198 
($\alpha$-enhanced chemical mixture) are taken into account.

\item For each metallicity, either two or three luminosity levels are included for fundamental (FU) 
or first overtone (FO) pulsators, to take into account evolved RRLs that are brighter than the ZAHB. 
The three possible luminosity levels are $\log{L_{ZAHB}}$ (A),  $\log{L_{ZAHB}}$ + 0.1 (B), 
and $\log{L_{ZAHB}}$ + 0.2 with mass 10\% lower than the other models (C).

\item The effective temperatures range between 7200 (bluest model on the 
ZAHB) to 5300 K (reddest model of the brightest luminosity level) with steps
of 100 K. The individual A, B, and C sequences include from four to 
eleven FU models and from two to seven FO models. This range covers the 
temperature width of the instability strip. 
\end{enumerate}

In \citet{marconi2015} we transformed the bolometric light curves of the quoted grid 
into the 2MASS JHK and the UBVRI photometric bands using the bolometric 
corrections and color-temperature relations obtained from the synthetic 
spectra provided by \citet{CastelliKurucz2003}. 
To transform the same models into the IRAC\footnote{Transmission curves for IRAC available at: \url{http://irsa.ipac.caltech.edu/data/SPITZER/docs/irac/calibrationfiles/spectralresponse/}}
 and the \emph{WISE}\footnote{Transmission curves for WISE available at: \url{http://wise2.ipac.caltech.edu/docs/release/prelim/expsup/sec4\_3g.html}}
 photometric bands,
we decided to perform a test to estimate the impact that different sets 
of synthetic spectra have on MIR bands. In particular, we used the 
atmosphere models provided by \citet{BrottHauschildt2005} 
(covering $2700~\mathrm{K} < T_{eff} < 10,000$~K and $-0.5 \leq \log g \leq 0.5$)
and by \citet{CastelliKurucz2003} (covering $3500~\mathrm{K} < T_{eff} < 50,000$~K and $0.0 \leq \log g \leq 5.0$).
%
The adopted synthetic spectra are available for a wide range of [Fe/H], namely for $-4.0 < $~[Fe/H]~$ 
< +0.5$. The \citet{BrottHauschildt2005} spectra are also available for several [$\alpha$/Fe] values in 
[-0.2, +0.8] with a spacing of 0.2 dex, while the \citet{CastelliKurucz2003}  ones are computed only for 
[$\alpha$/Fe]=0.0 and +0.4. For present calculations we adopted the spectra with [$\alpha$/Fe]=$+0.4$ 
in both cases. Since the two spectral libraries adopt different assumptions 
concerning the solar abundances (the \citet{BrottHauschildt2005} adopted the \citet{GrevesseNoels1993} solar
abundances whereas the \citet{CastelliKurucz2003} the ones from \citet{GrevesseSauval1998}), we placed them
on the same scale by converting the [Fe/H] of each
synthetic spectrum into the total metallicity $Z$. 
The relation used for this conversion is $\log (Z/X) - \log(Z/X)\sun = [Fe/H] + 0.35$ where
$\log (Z/X)\sun = -1.61$ \citep{pietrinferni2006}.

\begin{figure*}[tbp]
\centering
\includegraphics[scale=0.5]{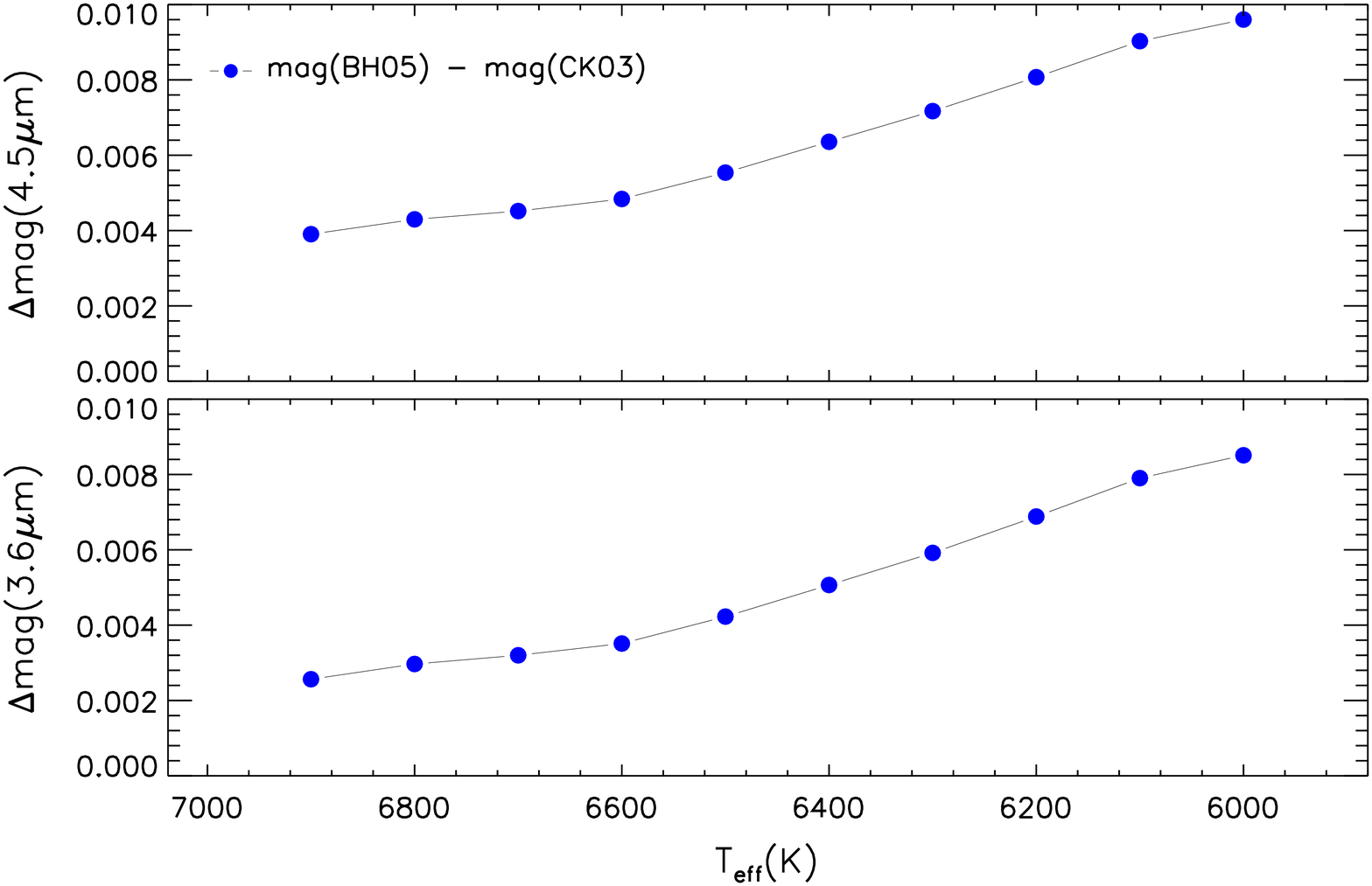}
\caption{ Top: Difference in [4.5] magnitudes as a function of the effective temperature for a set of FU models computed at fixed stellar mass 
(0.716 $M_\odot$), luminosity ($\log L/L_\odot =1.72$), and chemical composition ($Z=0.0003, Y=0.245$). Bolometric light curves predicted by 
hydrodynamical pulsation models were transformed into the observational plane by using color-temperature relations provided by \citet{CastelliKurucz2003} and by \citet{BrottHauschildt2005}. $T_{eff}$ spans the range between the blue and red edge of the instability strip. Bottom: Same as the top, but for the [3.6] magnitudes.}
\label{fig:CTs}
\end{figure*}

We obtained the bolometric corrections (BC) for all the $T_\mathrm{eff}$, 
$\log\,g$ and [Fe/H] ($Z$) values available in the spectral library. Then, we computed the 
magnitudes by interpolating the BC tables at the requested $T_\mathrm{eff}$, $\log\,g$ and 
total metallicity $Z$ along each bolometric light curve. The transformations were computed 
for the two different sets of atmosphere models and we found that the difference is negligible. 
Data plotted in Figure~\ref{fig:CTs} display that the difference in MIR mean magnitudes 
ranges from a few thousandths close to the blue (hot) edge of the instability strip to 
at most one hundredth of a magnitude close to the red (cool) edge of the instability 
strip. Data plotted in this figure refer to a sequence of FU models constructed at 
fixed mass, luminosity, and chemical composition (see labeled values). However, 
the difference is minimal over the entire grid of models. For the above reasons and for 
homogeneity with previous predictions, we decided to use the BC and the CT relations 
provided  by \citet{CastelliKurucz2003}.

\begin{figure*}[tbp]
\centering
\includegraphics[scale=0.8]{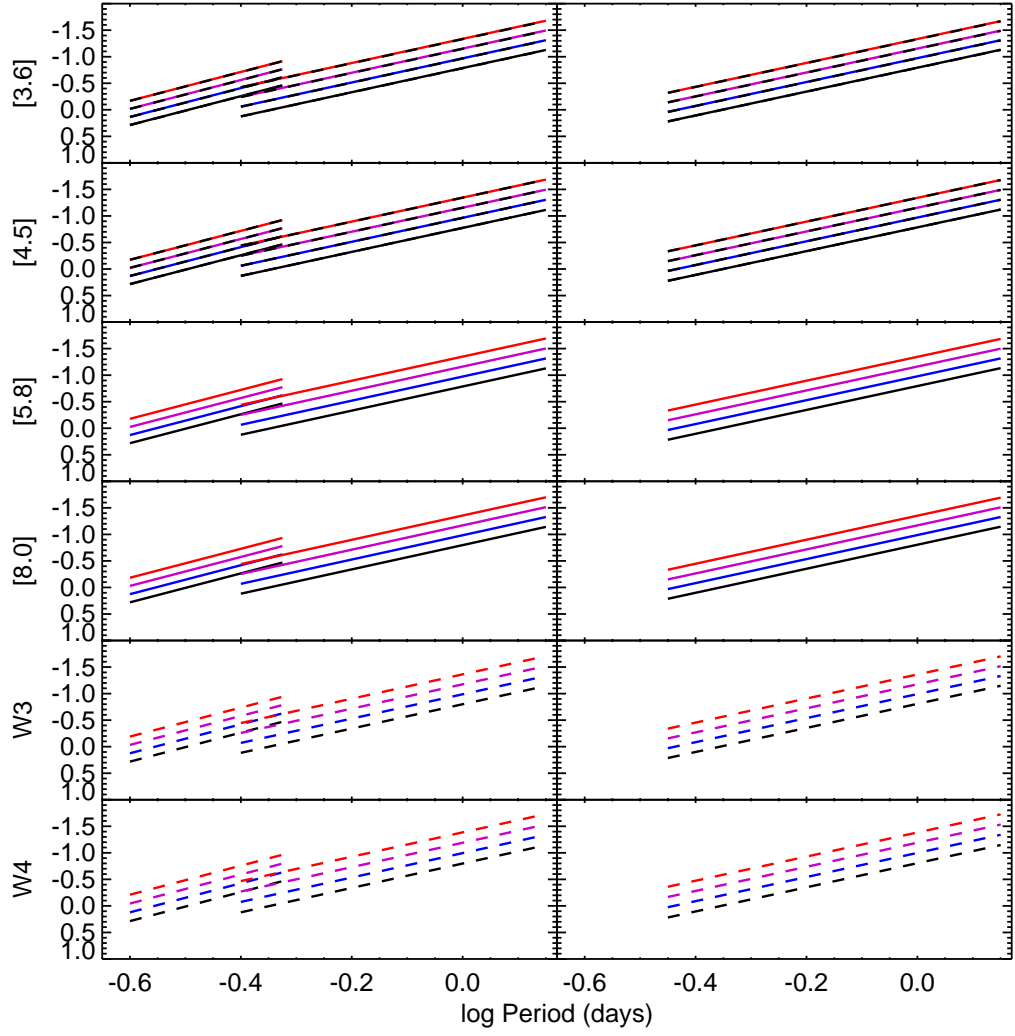}
\caption{Predicted RRL period-luminosity-metallicty relations for IRAC (solid lines) and \emph{WISE} (dashed lines) bands. Left panels show separate RRc and RRab relations, while the right panels show the fundamentalized relations. Four metallicities are plotted: [Fe/H] $=0$ (bottom black line), [Fe/H]$=-1.0$ (blue line), [Fe/H]$=-2.0$ (purple line), and [Fe/H] $=-3.0$ (top red line). }
\label{fig:theoryPLs}
\end{figure*}

Finally, we have fitted the periods, mean magnitudes and metallicities---now transformed into iron 
abundances---and obtained the coefficients of the PLZ relations. The mean magnitudes for the entire grid of 
models are given in Tables~\ref{tab:fo-means} and \ref{tab:fu-means} for the FO and FU pulsators 
respectively. The coefficients are given in Table~\ref{tab:PLZ}, and the relations for four metallicities 
([Fe/H] from 0 to -3.0 dex) are plotted in Figure~\ref{fig:theoryPLs}. Preliminary results based on two 
different metal abundances and three different helium contents indicate that pulsation properties of RR Lyrae
are minimally affected by the helium content. The key variation between models with canonical and enhanced 
helium content is the luminosity, and in turn the pulsation period \citep{marconi2011}. A more detailed
investigation will be addressed in a forthcoming paper (Marconi et al. 2017, in preparation), where the 
entire grid of model light curves, from the optical to MIR bands and covering the entire range in metallicity and 
helium abundance of RRL, will also be published.

We also obtained the coefficients 
for two and three band period-Wesenheit-metallicity (PWZ) relations, where the Wesenheit magnitude 
is defined as $W(B_1,B_2) = M_{B_1} - \alpha(M_{B_1}-M_{B_2})$ or $W(B_1,B_2,B_3)=M_{B_1}-\alpha
(M_{B_2}-M_{B_3})$. The coefficients of the color term, $\alpha$, have been fixed according to the 
reddening law we have adopted \citep{cardelli1989}. The coefficients for the color term and 
the PWZ relations are given in Tables~\ref{tab:PWZ2} and \ref{tab:PWZ3} for two and three bands
PWZ. For readers interested in the coefficients of the PWZ relation assuming a different reddening law, we provide 
a Python program\footnote{Available at \url{https://github.com/jrneeley/Theoretical-PWZ-Relations}}, in which you can input the coefficients ($A_\lambda/A_V$) of an alternative reddening law, and it will
output the corresponding coefficients of the PWZ relation. The full versions of Tables~\ref{tab:fo-means}, \ref{tab:fu-means}, \ref{tab:PWZ2}, and \ref{tab:PWZ3} are available in the online journal and portions are shown here for form 
and content.

By comparing the results described above with those obtained by \citet{marconi2015} in 
their Table~6, we outline that
{\it i)} The dispersion of the relations is constant (within 0.01 mag)
in the wavelength range from the $H$ to the $W4$ band, with values 
of the order of $\sigma_{FO}$=0.02 mag, $\sigma_{FU}$=0.03-0.04 mag 
and $\sigma_{FO+FU}$=0.04 mag.
{\it ii)} The metallicity coefficient shows no significant wavelength dependence from the $R$ to 
the IRAC bands, and is mainly due to the change in luminosity. This is because the metallicity 
dependence of the BCs effectively cancels out the change in $T_\mathrm{eff}$, resulting in a minimal
change in color ($\sim 0.01$ mag or less over the entire [Fe/H] range) as you increase metallicity.

\section{Optical, NIR, and MIR data}\label{obs}

To test the model PLZ relations described in Section~\ref{chapt_intro}, we have compiled multi-wavelength observations for a sample of 55 nearby Galactic RRL. The sample was based on stars included in the \emph{Hipparcos} catalog, limited to stars with $V < 11$ mag, $A_V < 0.5$, and an expected final \emph{Gaia} parallax with 2-3\% accuracy. Most of the observations were collected as part of the Carnegie RR Lyrae Program (CRRP, PID 90002), and were published in \citet{monson2016}. The pulsation period, metallicity, extinction, and available distance moduli of the stars in our sample are listed in Table~\ref{tab:stars}. In addition to the Galactic sample, we compiled observations of RRL in the globular cluster M4. All data has been homogenized to the following photometric system: Kron-Cousins RI, 2MASS $JHK_s$, \emph{Spitzer} S19.2 3.6 and 4.5 $\micron$, \emph{Spitzer} S18.25 5.8 and 8.0 $\micron$, and \emph{WISE} $W1$, $W2$ and $W3$ \citep[see][for details]{monson2016}. 



\subsection{Galactic RRL}

The $RIJHK$ data set was built from three sources: \citet{monson2016}, \citet{klein2014}\footnote{Available online at: \url{http://w.astro.berkeley.edu/~cklein/dissertation.pdf}}, and \citet{feast2008}. The data set from \citet{monson2016} combines observations collected for CRRP with the Three-Hundred Millimeter telescope with archival observations from individual studies (and places them on a homogeneous photometric system), and was used as the basis for our sample of Galactic RRL. For stars without multi-epoch observations in this data set, we supplemented with the multi-epoch observations available in \citet{klein2014}.  Light curves from \citet{klein2014} were obtained with the Nickel 1 meter telescope at Lick Observatory and 1.3 meter PAIRITEL telescope at the Fred Lawrence Whipple Observatory and were fit with Fourier series. Finally, \citet{feast2008} presented a procedure to estimate mean magnitudes from single epoch 2MASS JHK measurements of RRL. For any stars still missing multi-epoch $JHK$ observations from the previous two data sets, we adopted the estimated mean magnitude from \citet{feast2008}. The \citet{feast2008} templates are not very precise, and we derived a single uncertainty for each band according to the standard deviation of the residuals between the template mean magnitude and derived mean magnitude of stars for which we have well sampled light curves. This results in uncertainties in the adopted mean magnitudes of 0.06, 0.04, and 0.07 mag for $JHK$, respectively. 

\begin{figure*}[tbp]
\centering
\includegraphics[scale=0.9]{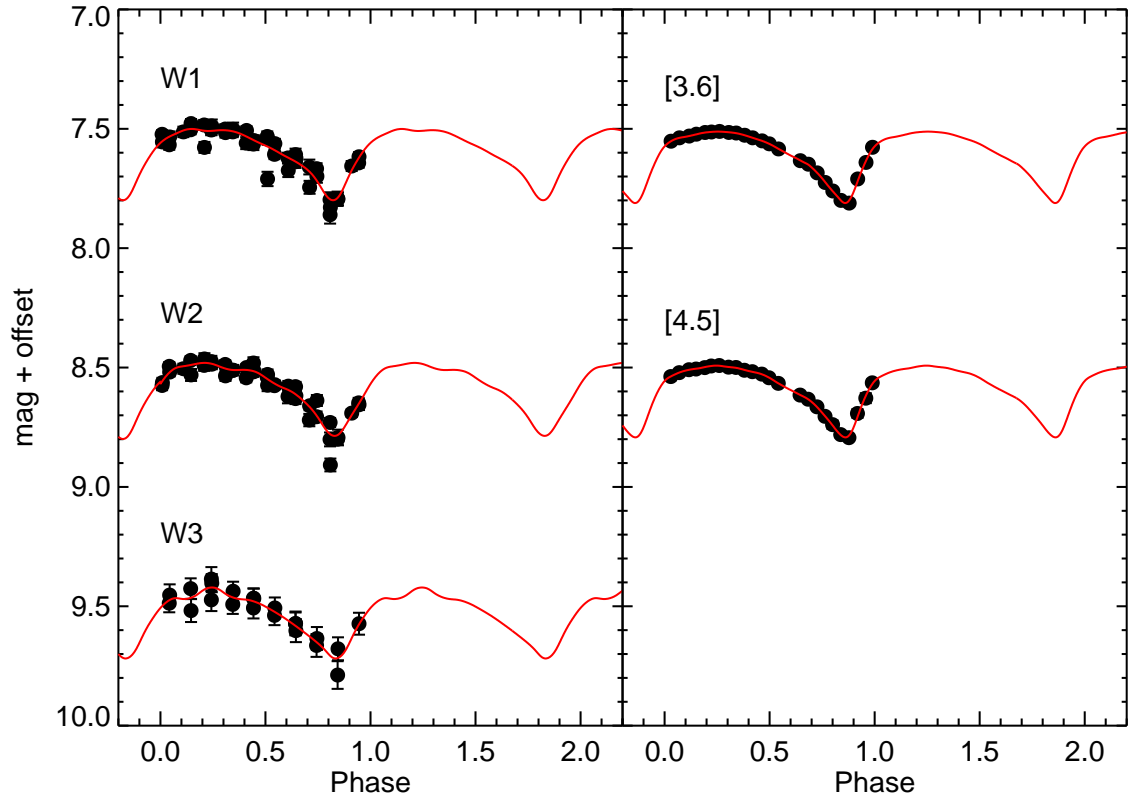}
\caption{Sample WISE and IRAC light curves for the Galactic RRL star, SU Dra. The W1, [3.6] and W3 bands are offset by $ \pm 1$ mag for better visibility. The smoothed light curve generated by GLOESS is shown as the red line.   }
\label{fig:lcvs}
\end{figure*}

Mid-infrared observations for our sample are also presented in \citet{monson2016}, that combined the \emph{Spitzer} 3.6 and 4.5 $\micron$ observations obtained for CRRP with \emph{WISE} photometry. The first two bands in the \emph{WISE} photometric system are very similar to the IRAC bands, but we have elected to keep the \emph{WISE} and IRAC photometry separate in this paper for two main reasons. First, for most of the stars in our sample, adding in \emph{WISE} photometry degrades the quality of our IRAC light curves (Figure~\ref{fig:lcvs} compares the quality of \emph{WISE} and IRAC light curves side by side). Second, the \emph{WISE} and IRAC passbands are not identical, and it is unclear if there is an offset between the two calibrations. When comparing the average magnitudes of RRL obtained with \emph{Spitzer} and \emph{WISE}, \citet{monson2016} saw a small offset. However, the AllWISE explanatory supplement\footnote{Available online at: \url{http://wise2.ipac.caltech.edu/docs/release/allwise/expsup/}} ~found no offset with the first two IRAC bands. Given this unresolved discrepance, we re-derived the average magnitudes separately for the IRAC and \emph{WISE} data presented in \citet{monson2016}. The IRAC light curves are covered by a minimum of 24 epochs over a single pulsation cycle (some stars were observed with up to 134 epochs to fill gaps in \emph{Spitzer's} schedule). We note that the Spitzer Science Center (SSC) recently released the final calibration of warm mission \emph{Spitzer} data (S19.2). The new calibration included new flux conversions, linearity solution for the 3.6 $\micron$ band, and flat fields, and is now consistent with the final cryogenic mission calibration (S18.25) defined by \citet{carey2012}. The \emph{WISE} photometry comes from the AllWISE data release, and provides an average of 36, 36, and 18 random epochs in the $W1$, $W2$, and $W3$ bands, respectively.  

The GLOESS method \citep[and references therein]{neeley2015, monson2016} was used to fit smoothed light curves and derive the mean intensity magnitude in each of the bands. For stars in common with the CRRP sample, we found no significant difference in the mean magnitudes fit with fourier series \citep{klein2014} and the GLOESS method \citep{monson2016}. The final count for each band is as follows: 32 in R, 55 in I, 54 in J, 53 in H, 54 in K, 51 in [3.6], 51 in [4.5], 55 in W1, 55 in W2, and 54 in W3.

\subsection{M$4$ RRL}

Photometry for the RRL in the globular cluster M4 is presented in \citet{stetson2014} ($RIJHK$) and in \citet{neeley2015} (\emph{Spitzer} [3.6] and [4.5]). The $RIJHK$ data were assembled from $\sim65$ data sets mostly taken at the VLT. The IRAC 3.6 and 4.5~$\micron$ photometry presented in \citet{neeley2015} was collected as part of CRRP, and has been updated in this paper to the most recent \emph{Spitzer} calibration (S19.2) using the same calibration procedure as for the Galactic RRL. All photometry was undertaken using the DAOPHOT/ALLSTAR/ALLFRAME \citep{daophot, daophot2, allframe} suite of programs.

For this work, we have also performed new photometry of single-epoch archival observations of M4 from \emph{Spitzer's} cryogenic mission 5.8 and 8.0 $\micron$ bands. The cluster was observed with a medium scale 11-point cycling dither pattern, with a frame time of 100 seconds. Aperture photometry was performed on Basic Calibrated Data (BCDs) generated by the S18.25 pipeline with a 3-3-7 pixel aperture, and calibrated to the photometric system defined by \citet{carey2012}. The aperture corrections used were 1.1356 and 1.2255 at 5.8 and 8.0~$\micron$ respectively. Photometry at the 11 dither positions was averaged to obtain a single measurement. These observations were obtained 8 years before the CRRP \emph{Spitzer} observations, and we were unable to reliably determine the pulsation phase and construct a template to estimate the mean magnitude. Therefore we elected to use the single epoch observation as the estimated mean magnitude with an uncertainty equal to half the amplitude in the 3.6 or 4.5 $\micron$ bands. These results as well as the updated mean magnitudes from \citet{neeley2015} are available in Table~\ref{tab:M4data}.

\section{Empirical PL relations}\label{empiricalPL}

We can directly compare our synthetic PLZ relations with some recent empirical relations. Since many empirical measurements have been made on single metallicity populations (i.e. in globular clusters), the metallicity component has not been reliably measured and only PL relations are available. One exception to this is the empirical PLZ relation in the $W1$ and $W2$ bands measured using several globular clusters \citep{dambis2014}. Both the period slope and the zero point are consistent between the theoretical and empirical methods. However the metallicity term they measure (0.096 mag/dex in $W1$ and 0.108 mag/dex in $W2$) is smaller than in our synthetic PLZ relations by about $4\sigma$, but we note that their method is reliant on the accuracy of the visual band magnitude - metallicity relation. \citet{madore2013} also measured PL relations in the \emph{WISE} bands, but using only four RRab stars with parallax measurements. Due to their limited sample, the uncertainties on their derived parameters are very large ($\pm 0.9$ mag in slope, $\pm 0.2$ mag in the zero point), and they were not able to see a metallicity dependence (these four stars also cover an extremely limited range in [Fe/H]).

We have recently completed a multi-wavelength analysis of the PL relation for the globular cluster M4 (NGC 6121). The $RIJHK$ relations were presented in \citet{braga2015} and IRAC 3.6 and 4.5 $\micron$ relations were originally presented in \citet{neeley2015}. For consistency, we have updated the calibrated relations from \citet{neeley2015} to the new \emph{Spitzer} S19.2 calibration, as well as extended the relations to the 5.8 and 8.0 $\micron$ bands. The new first overtone, fundamental, and fundamentalized relations are given in Table~\ref{tab:M4PLs}. For the 3.6 and 4.5 $\micron$ bands, the slope was fixed using the cluster RRL, and the zero point was determined using five Galactic RRL (RR Lyr, RZ Cep, SU Dra, UV Oct, and XZ Cyg) with geometric parallaxes measured using the $HST$ Fine Guidance Sensor \citep{benedict2011}. The $\sigma$ quoted is the dispersion of the cluster RRL around the PL relation. The fundamentalized calibrated relations result in a distance modulus of $\mu_0 = 11.353\pm0.095$ at 3.6 $\micron$ and $\mu_0=11.363\pm0.095$ at 4.5 $\micron$, assuming extinction values calculated specifically for M4 ($A_V = 1.39 \pm 0.01$ and $R_V=3.62$) in \citet{hendricks2012}. For the longer 5.8 and 8.0 $\micron$ bands, the slope is based on single epoch photometry of the M4 RRL, so the uncertainties and dispersion are significantly larger. In addition, no photometry of the five calibrating RRL was available in these bands, so the calibrated zero point is determined by applying the average distance modulus measured from the 3.6 and 4.5 $\micron$ bands to the apparent zero point. 

\begin{figure*}[tbp]
\centering
\includegraphics{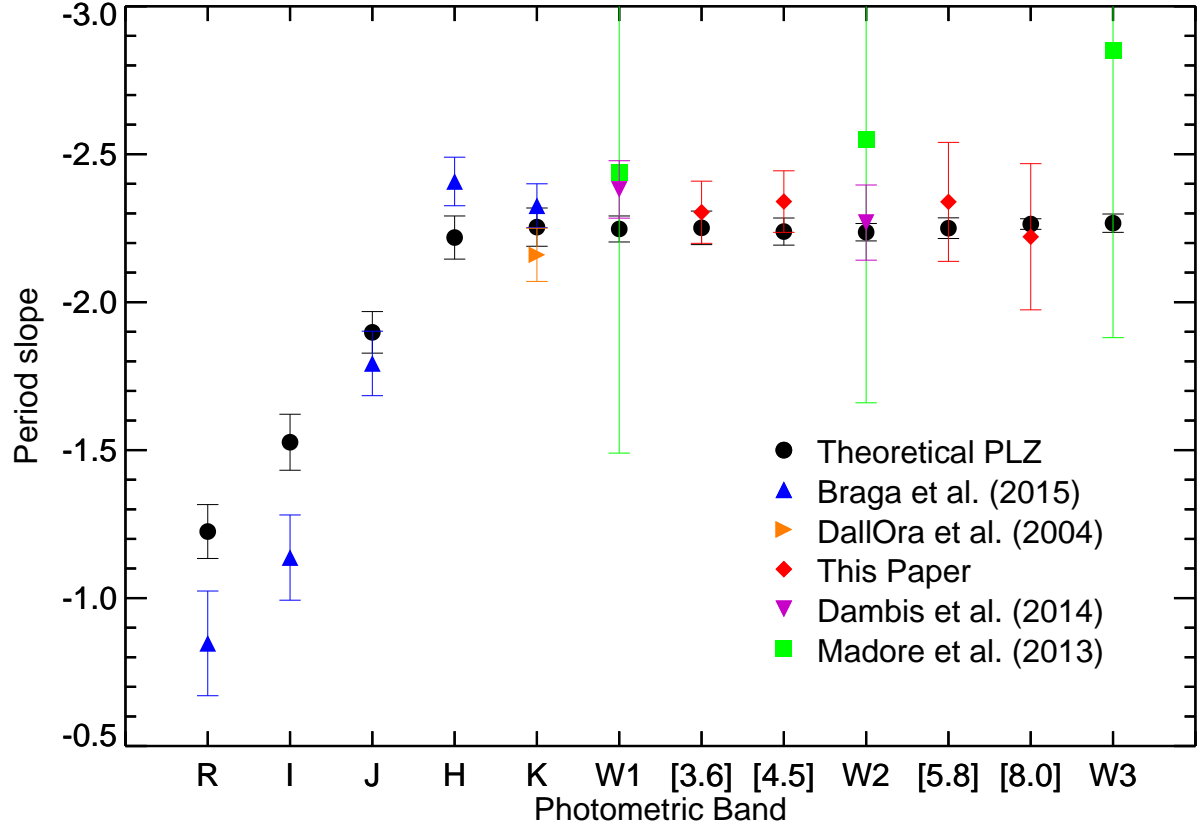}
\caption{The period dependence ($b_\lambda$) for different filters. The predicted coefficients of PLZ relations from \citet{marconi2015} and Table~\ref{tab:PLZ} of this work are shown as black circles. Empirical measurements from the literature are shown for comparison according to the legend. }
\label{slope}
\end{figure*}

A comparison between the period dependence derived from theoretical methods and the empirical methods described above is shown in Figure~\ref{slope}. The theoretical period dependence is shown as filled circles, and empirically derived slopes are over-plotted according to the legend. Overall the predicted period dependence in all bands is consistent with results from the literature, and we do observe the predicted flattening of the period dependence at longer wavelengths. The observed slope of the PL relation in M4 is within $1\sigma$ of the predicted slope in the $J$, $K$, and IRAC bands. In the $R$, $I$ and $H$ bands, the slope of the observed PL relations differs from the predicted relation by 2, 2.7, and 2$\sigma$ respectively.  This discrepancy could be due to the larger scatter in the PL relation at shorter wavelengths, but we note that we find a better agreement with the synthetic HB - based slopes by \citet{catelan2004} in these bands. Ongoing work on more globular clusters in the CRRP sample will help to better characterize the period dependence in the RRL PLZ relation.

\section{Galactic RRL distances}\label{galactic}

In addition to simply comparing theoretical and empirical results, we can use our synthetic PLZ relations to fit the data and obtain predicted distance moduli and extinctions for all of the RRL in our sample. We used the synthetic PLZ relations to obtain predicted absolute magnitudes, $M_{\lambda}$, in the $RIJHK$, IRAC, and \emph{WISE} bands. From these predicted absolute magnitudes, reddened distance moduli ($m_\lambda - M_\lambda$) were calculated, and then used in combination with a universal reddening law to fit both the true distance modulus ($\mu_0$) and visual band extinction ($A_V$) of each star using a weighted least squares fit of the form
\begin{equation}
m_{\lambda} - M_{\lambda} = \mu_{0} + A_{V}  (A_{\lambda}/A_{V})
\end{equation}
where $A_\lambda/A_V$ are the coefficients of the reddening law. This technique was first introduced by \citet{freedman1985,freedman1991} and has recently been employed on Classical Cepheids by \citet{gieren2005} and \citet{inno2016}. The reddening law we adopted was that of \citet{cardelli1989}, which we extended into the IRAC and \emph{WISE} bands according to \citet{indebetouw2005} and \citet{madore2013} respectively. The weights in the fit were determined by the sum in quadrature of the uncertainties in the apparent mean magnitudes and predicted absolute magnitudes of the star, where the uncertainty of the predicted absolute magnitude is given by the dispersion of the PLZ relation of the corresponding wavelength. As a consequence, the IRAC bands have the most weight in determining the distance modulus, since the PLZ relations have the lowest dispersion at these wavelengths, and the IRAC mean magnitudes are more precise than those determined with \emph{WISE} data. The shorter wavelengths provide the leverage necessary to fit the extinction. The results of this fit are given in the last two columns of Table~\ref{tab:stars}. 

\begin{figure*}[tbp]
\centering
\includegraphics[scale=1]{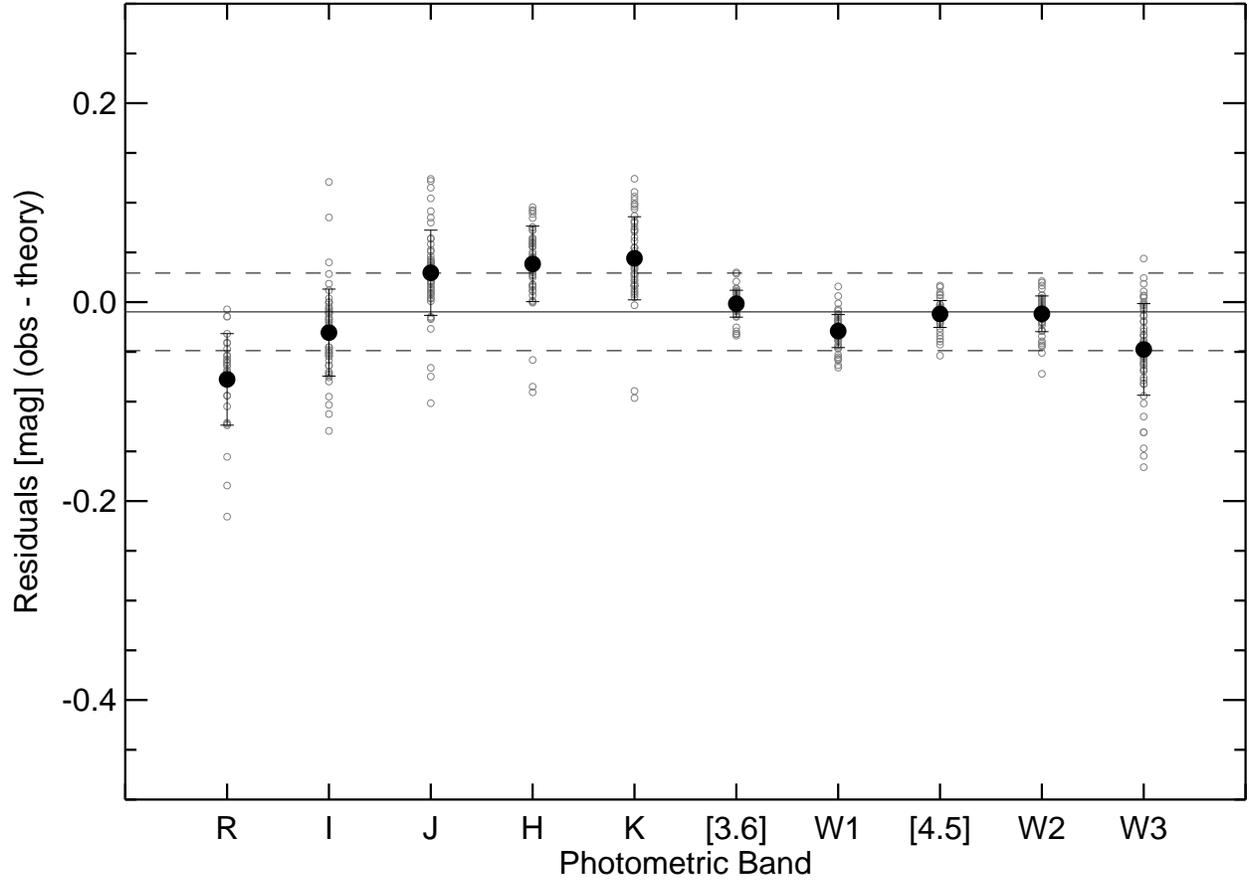}
\caption{Residual between the absolute magnitude using the best fit distance moduli and extinctions and the predicted absolute magnitude from theoretical PLZ relations for all galactic RRL in each band. Individual stars are shown with open circles, and the average residual for each band is shown as a filled circle. The error bars represent the standard deviation in the residuals of the individual stars. The solid line is the weighted mean of the average residuals, and the dashed lines are $1\sigma$ from the mean. Note that because this is a weighted fit, the solid line passes the points with the most weight ([3.6] and [4.5]) and not through zero. }
\label{fig:galresiduals}
\end{figure*}

To assess the ability of the theoretical PLZ relation to fit the distance moduli of all sample stars at different wavelengths, we have transformed all average magnitudes into semi-empirical absolute magnitudes using the best fit distance moduli and extinctions from Table~\ref{tab:stars}. The residuals between the calculated (from observed apparent magnitude) and predicted (from the PLZ relation) absolute magnitudes in each band are shown in Figure~\ref{fig:galresiduals}. Individual stars are shown as small open circles, while the average residual in each band is over-plotted as filled circles. The dashed lines represent the standard deviation around the mean of the average residuals. The IRAC, W1, and W2 bands offer the smallest dispersion. Predicted absolute magnitudes in the $JHK$ bands tend to be brighter than the calculated absolute magnitude. There are three possibilities for this discrepancy: 1) the reddening law is inappropriate for these bands, 2) mean magnitudes calculated from $JHK$ templates are systematically too faint, or 3) the color temperature relations used to transform the bolometric light curves into the observational plane are incorrect for NIR wavelengths. The first point is unlikely, because the effect of reddening is low at NIR wavelengths, and the coefficients of the reddening law would have to change by as much as 50\%. We see no systematic offset between mean magnitudes derived from $JHK$ templates and observed light curves, but we note that less than half the stars in our sample have NIR light curves with high enough phase coverage to measure an accurate mean magnitude. Further observations are needed to see if this resolved the discrepancy with theory, or if the color temperature relations need some adjustment in the $JHK$ bands.

\begin{figure*}[tbp]
\centering
\includegraphics[scale=0.8]{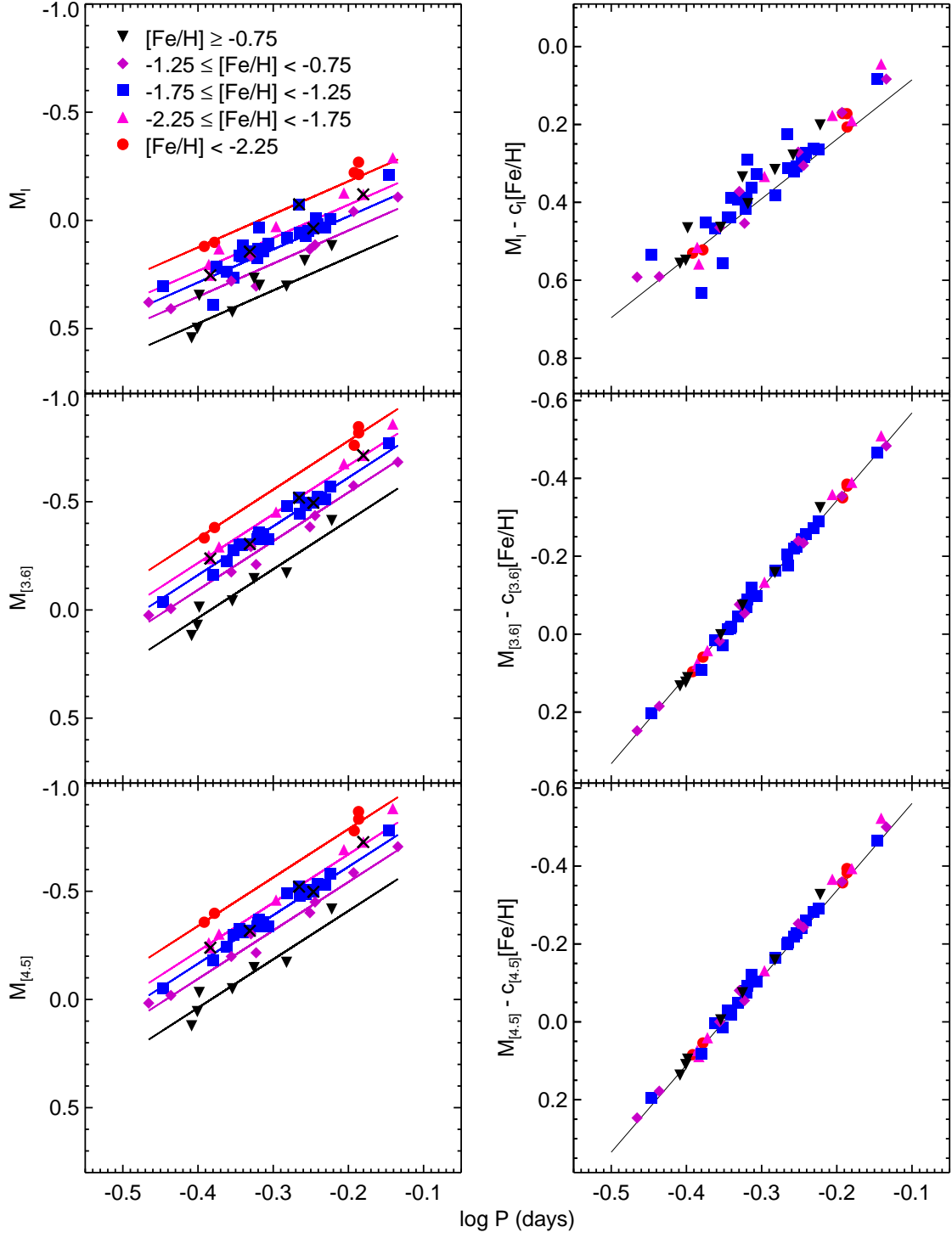}
\caption{Calculated absolute magnitude versus period of field RRL stars for the I, 3.6, and 4.5 $\micron$ bands. Left panels -- The stars have been separated into metallicity bins from low (black triangles) to high (red circles) metal abundance. Predicted relations for the median of each metallicity bin are shown with solid lines. Right panels -- The metallicity term ($c_{\lambda}*[Fe/H]$) has been subtracted from the vertical axis (normalized to solar metallicity, [Fe/H]=0.0 dex.}
\label{fig:PLZ}
\end{figure*}

\begin{figure*}[tbp]
\centering
\includegraphics[scale=0.8]{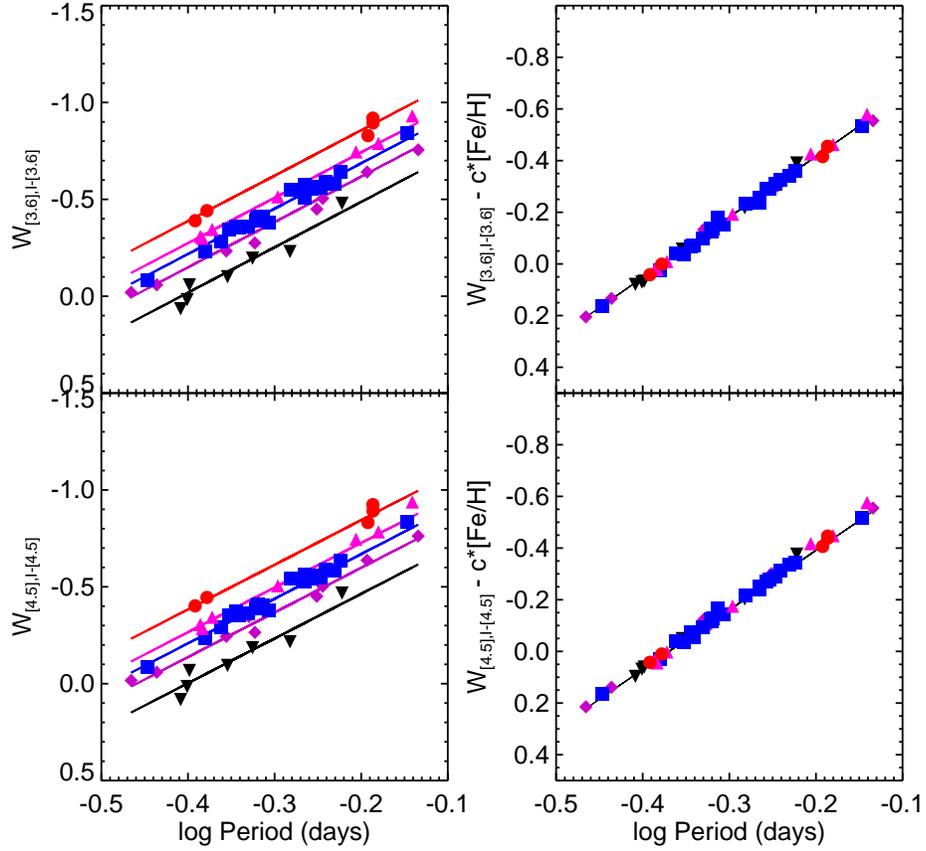}
\caption{Period-Wesenheit-metallicity relation of field RRL stars. Predicted relations for the five metallicity bins are shown with solid lines. In the right panel the metallicity term has been subtracted form the vertical axis. }
\label{fig:PWZ}
\end{figure*}

In \citet{klein2014}, a multi-wavelength approach similar to ours was used to fit distances and extinctions of RRL, but the effects of metallicity were ignored. In their Figure 6.20, PL relations from optical to MIR are shown with remarkably small scatter. However the scatter is by design artificially small, since the optimal solution is estimated minimizing the residuals of the different bands. Our synthetic PLZ relations provide a method to estimate the effect of metallicity on the determination of distances. The left panels of Figure~\ref{fig:PLZ} show how the RRL in our sample lie in the period-magnitude plane when metallicity is ignored. Stars were divided into five metallicity bins, and the predicted PL relation for the median of each bin is shown as the solid line in the corresponding color. The dispersion around individual relations is quite low (with much of it still due to a spread in metallicity), but if you consider all RRL covering a large range of metallicity as a single population, the dispersion around the average metallicity is large (0.12, 0.13, and 0.13 in the $I$, [3.6], and [4.5] bands respectively). In the right panels of Figure~\ref{fig:PLZ}, the metallicity term has been subtracted out, and the vertical axis is now $M_{\lambda} - c_{\lambda}[Fe/H]$. Now the dispersion around the theoretical relation is almost an order of magnitude smaller (0.053, 0.019, and 0.017 mag in the $I$, [3.6], and [4.5] bands respectively). These panels illustrate the drastic improvement you can expect in the accuracy of distance determinations when accounting for metallicity. Assuming the theoretical PLZ relations correctly characterize the effect on the zero point due to metallicity, using PLZ relations over PL relations will reduce the uncertainty of distance measurements from 13\% to 2\%. A similar argument can be shown using PWZ relations (Figure~\ref{fig:PWZ}). Here, the dispersion in the PW relation once metallicity has been removed is even smaller (0.012 mag for $I$-IRAC PWZ relations). This could be because PWZ relations are reddening free (they depend only on the reddening law, not the extinction of individual objects), or because the PWZ relations include a color term and reduce the scatter because they take into account the width in temperature of the instability strip.

\subsection{Comparison with HST}

Prior to the \emph{Gaia} release, the only geometric measurement of RRL distances was the parallax of five stars (RR Lyr, RZ Cep, SU Dra, UV Oct, and XZ Cyg) measured using \emph{HST} Fine Guidance Sensor \citep{benedict2011}. With only five calibrators available, the precision of the absolute zero point of RRL PL relations has been severely limited. In this section, we compare the \emph{HST} parallaxes with results based on the theoretical PLZ relations. 

\begin{figure*}[tbp]
\centering
\includegraphics{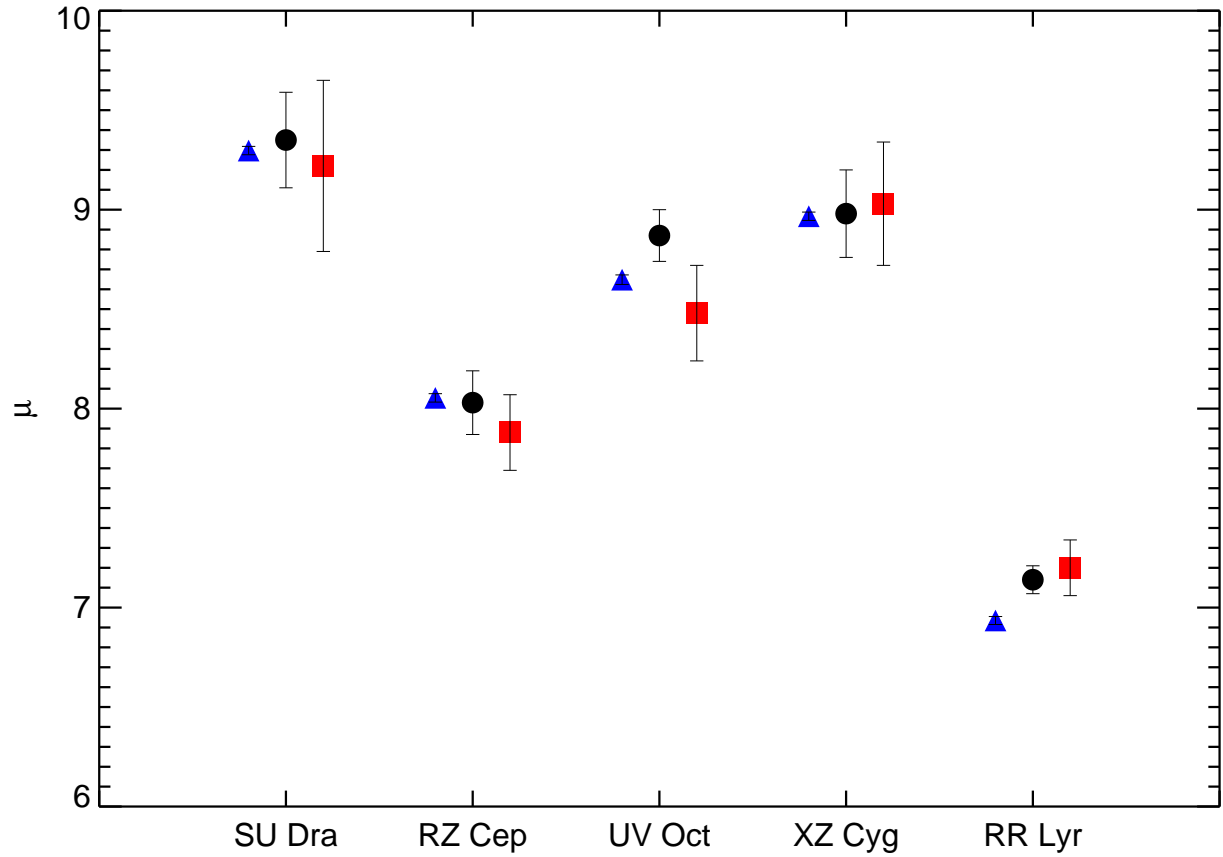}
\caption{A direct comparison of the distance modulus of five stars obtained with three methods: 1) the method presented in this paper (blue triangles), 2) \emph{HST} parallax measurements from \citet{benedict2011} (black circles), and 3) proper motion parallaxes from TGAS (red squares). For both the \emph{HST} and TGAS distance moduli, the error bars are dominated by the error in the parallax. }
\label{fig:hst-dms}
\end{figure*}

\begin{figure*}[tbp]
\centering
\includegraphics[scale=0.8]{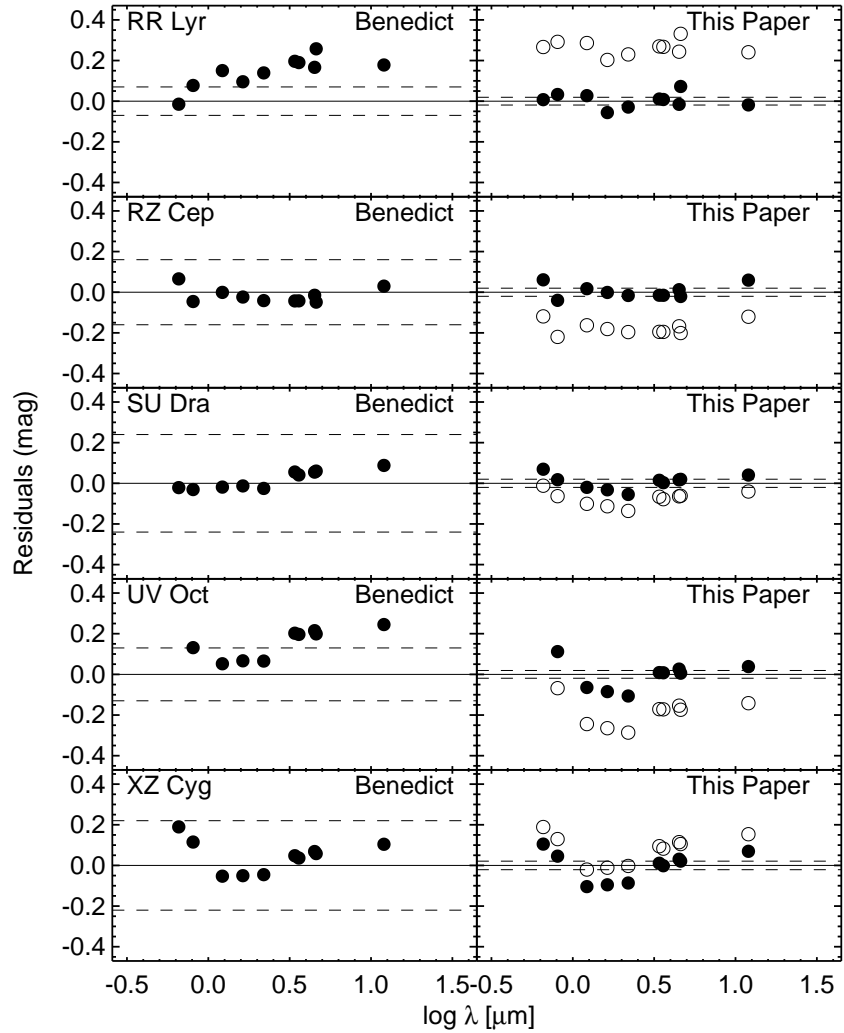}
\caption{Comparison of the offset with predicted absolute magnitude when using the HST derived distance modulus and extinction from \citet{benedict2011} (left panels) and in this paper (right panels) as a function of wavelength. The dashed lines indicate the uncertainty in the distance moduli. The right panels also include as open circles the residual when using the TGAS distances and the extinction from this paper. Trends with wavelength indicate the extinction is inconsistent with theoretical predictions, while on overall offset suggests the distance is inconsistent with theory.  }
\label{fig:Bencompare}
\end{figure*}

Table~\ref{tab:stars} presents the distance moduli and extinction derived in \citet{benedict2011}, compared with the value estimated in this work. The distance moduli comparison is also shown in Figure~\ref{fig:hst-dms}. Two stars, RR Lyr and UV Oct, exhibit a 2.7 and 1.6$\sigma$ tension, respectively, between our method and the HST distance moduli. Three stars (RR Lyr, SU Dra, and UV Oct) are also fit with a significantly larger extinction using our method. The extinctions adopted in \citet{benedict2011} are from \cite{feast2008}, where stars were fit using 3-D Galactic extinction models while assuming a distance based on $M_V -[Fe/H]$ or preliminary $PL(K)$ relations. Figure~\ref{fig:Bencompare} compares the residuals with predicted absolute magnitude when correcting observations using the HST parallaxes and extinctions (left panel) as a function of wavelength. The residuals of four stars (RR Lyr, RZ Cep, SU Dra, and UV Oct) present a trend with wavelength when using the HST derived parameters, indicating the extinction assumed for these stars in \citet{benedict2011} is incorrect. The vertical offset seen in RR Lyr and UV Oct is a consequence of the difference in distance modulus we measure. In contrast, the right panel shows no offsets or trends, suggesting good agreement with theory across all wavelengths. Clearly, the observations and theory are at odds for some of these stars. The discrepancy with predicted absolute magnitude shows no trend with period, distance, or metallicity, which indicates the problem may lie with individual parallax and extinction measurements. We should note that both RR Lyr and UV Oct exhibit the Blazhko effect, and it is possible that this is affecting the mean magnitude of these stars. However, XZ Cyg is also a Blazhko star, and all methods are in good agreement for this star. 

\subsection{Comparison with Gaia DR1}

\begin{figure*}[tbp]
\centering
\includegraphics{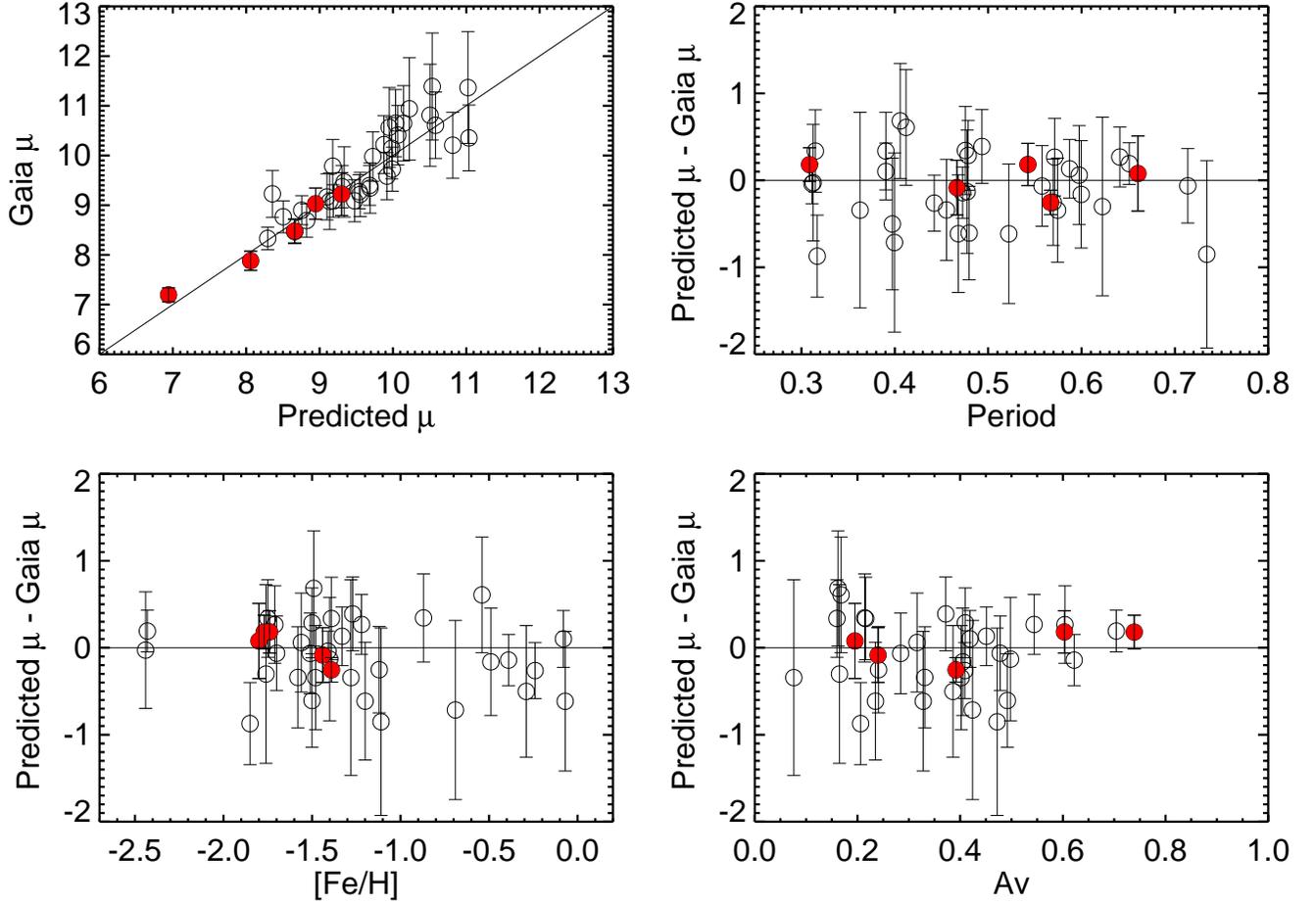}
\caption{\emph{Top left:} One-to-one comparison of the distance moduli derived in this paper and determined from the first release of TGAS parallaxes. The five stars with previously determined \emph{HST} parallaxes are highlighted with closed red circles. Given that all but two stars agree within $1\sigma$, this may indicate that the TGAS errors are overestimated as suggested in \citet{castertano2016} and \citet{gould2016}. \emph{Top right:} Residuals between predicted and TGAS parallaxes as a function of the pulsation period. \emph{Bottom left:} Residuals between predicted and TGAS parallaxes as a function of the metal abundance. \emph{Bottom right:} Residuals between predicted and TGAS parallaxes as a function of the fitted visual band extinction. }
\label{gaia}
\end{figure*}

We can also compare the distance moduli obtained above with the recent parallax measurements from the \emph{Gaia} mission. \emph{Gaia's} first data release (DR1) became public on September 14, 2016 \citep{gaia2, gaia1}. This release included proper motion parallax measurements for stars in common with the Tycho-2 catalog, and are based on the Tycho-Gaia astrometric solution \citep[TGAS,][]{michalik2015}. TGAS parallaxes were available for 41 of the stars in our sample, with uncertainties ranging from $\sigma_\pi/\pi=0.06$ to $\sigma_\pi/\pi=0.83$. The distance moduli from TGAS parallaxes are given in Table~\ref{tab:stars}. In Figure~\ref{gaia} we compare the TGAS results with the distance moduli obtained in this paper for stars with $<55\%$ uncertainty in the TGAS parallax. The top right panel is a one to one comparison between the two methods. The five stars with previously determined \emph{HST} parallaxes are highlighted with filled red circles. For the majority of stars, the methods agree within $1\sigma$. Only four stars (AM Tuc, RR Lyr, MT Tel, V Ind) are outliers, and all are within 2$\sigma$. Both the dispersion and individual uncertainty increases with distance. The dispersion is 0.30 for stars with $\mu_0 < 9.5$ and 0.46 for stars with $\mu_0 > 9.5$, and the average uncertainty in the TGAS distance is twice as large for stars with $\mu_0 > 9.5$ mag. The remaining panels in Figure~\ref{gaia} show the residual between the two methods as a function of stellar parameters, indicating there are no trends with period, metallicity, or $A_V$. This confirms that the residual error between our best fit value and the TGAS parallax are likely due to the statistical uncertainties in the TGAS results. 


\section{M4 distance}\label{M4}

We can use globular clusters as an additional laboratory to test the theoretical PLZ relations. Globular clusters offer the advantage that there are many RRL at roughly the same distance, reddening, and metallicity. For the closest globular cluster to us, M4 (NGC 6121), if we assume a 10pc radius the expected difference in distance modulus between stars in the front and back of the cluster is about 0.02 mag. Since this is smaller than the expected dispersion in the PLZ relation, we cannot accurately measure the distances and extinctions for individual stars. However, we can measure the average distance and extinction for the cluster as a whole.

\begin{figure*}[tbp]
\centering
\includegraphics[scale=0.8]{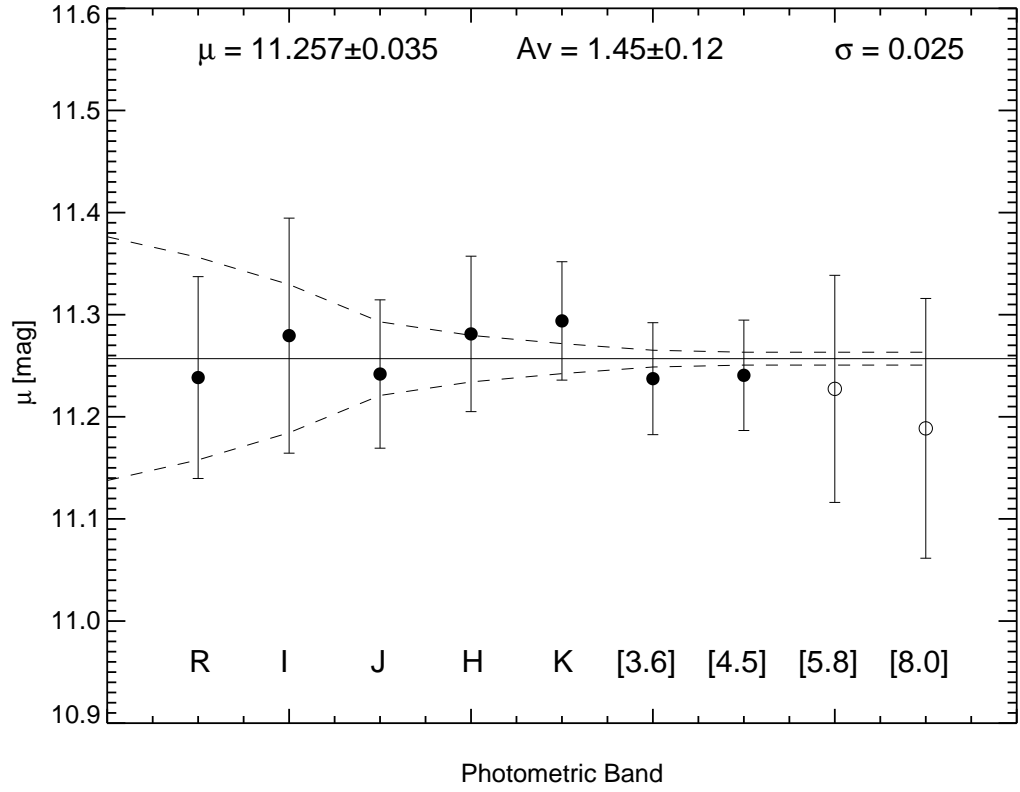}
\caption{Derived extinction corrected distance modulus of M4 for all available bands. Only points plotted with filled circles were used in calculating the band averaged distance modulus and extinction, and the remaining bands are shown only for reference. The solid line is the average distance modulus, and the dashed lines indicate how the uncertainty in extinction propagates with wavelength.  }
\label{fig:dms}
\end{figure*}

The distance modulus and reddening to M4 was fit in a similar method to the Galactic RRL, but instead of fitting an individual distance and extinction for each star, we derived the average $\mu_0$ and $A_V$. A reddened distance modulus for each band was calculated by averaging the difference between observations and predicted absolute magnitude from the theoretical PLZ relations. The true distance modulus and visual extinction were then fit using a least squares fit as in Equation 1, but now the reddening coefficients $A_{\lambda}/A_{V}$ are defined by a reddening law specific to M4 \citep{hendricks2012}. This reddening law accounts for the fact that the cluster is behind the $\rho$ Oph cloud by adopting a higher dust parameter ($R_V=3.62$) than the Cardelli law ($R_V = 3.1$). Figure~\ref{fig:dms} shows the extinction corrected distance moduli for each band. Single epoch archival data for the two longer \emph{Spitzer} bands are also shown for reference as open circles, but were not included in the fit. The dashed lines represent how the uncertainty in extinction propagates with wavelength. The final distance modulus is $\mu_0 = 11.257 \pm 0.035$, and the $V$ band extinction is $A_V = 1.45 \pm 0.12$. This is consistent with our results from purely empirical NIR and MIR PL relations, $\mu_0=11.283\pm0.010\pm0.018$ \citep{braga2015}, $\mu_0 = 11.353 \pm 0.095$ at 3.6 $\micron$ and $\mu_0=11.363\pm 0.095$ at 4.5 $\micron$ (Section~\ref{empiricalPL}) respectively. Our measured distance modulus of $\mu_0=11.257$ is also consistent with recent measurements from a variety of other primary distance methods, and we will highlight a few here \citep[see Section 2 of][for a more complete discussion]{braga2015}. \citet{hendricks2012} estimates $A_V = 1.39 \pm 0.01$ and $\mu_0 = 11.28 \pm 0.06$ so our measurements are within $1\sigma$. The distance to M4 has also recently been measured by \citet{kaluzny2013} using three eclipsing binary stars. They obtained $\mu_0 = 11.30 \pm 0.05$, which is also in agreement with our method. 

\begin{figure*}[tbp]
\centering
\includegraphics[scale=0.7]{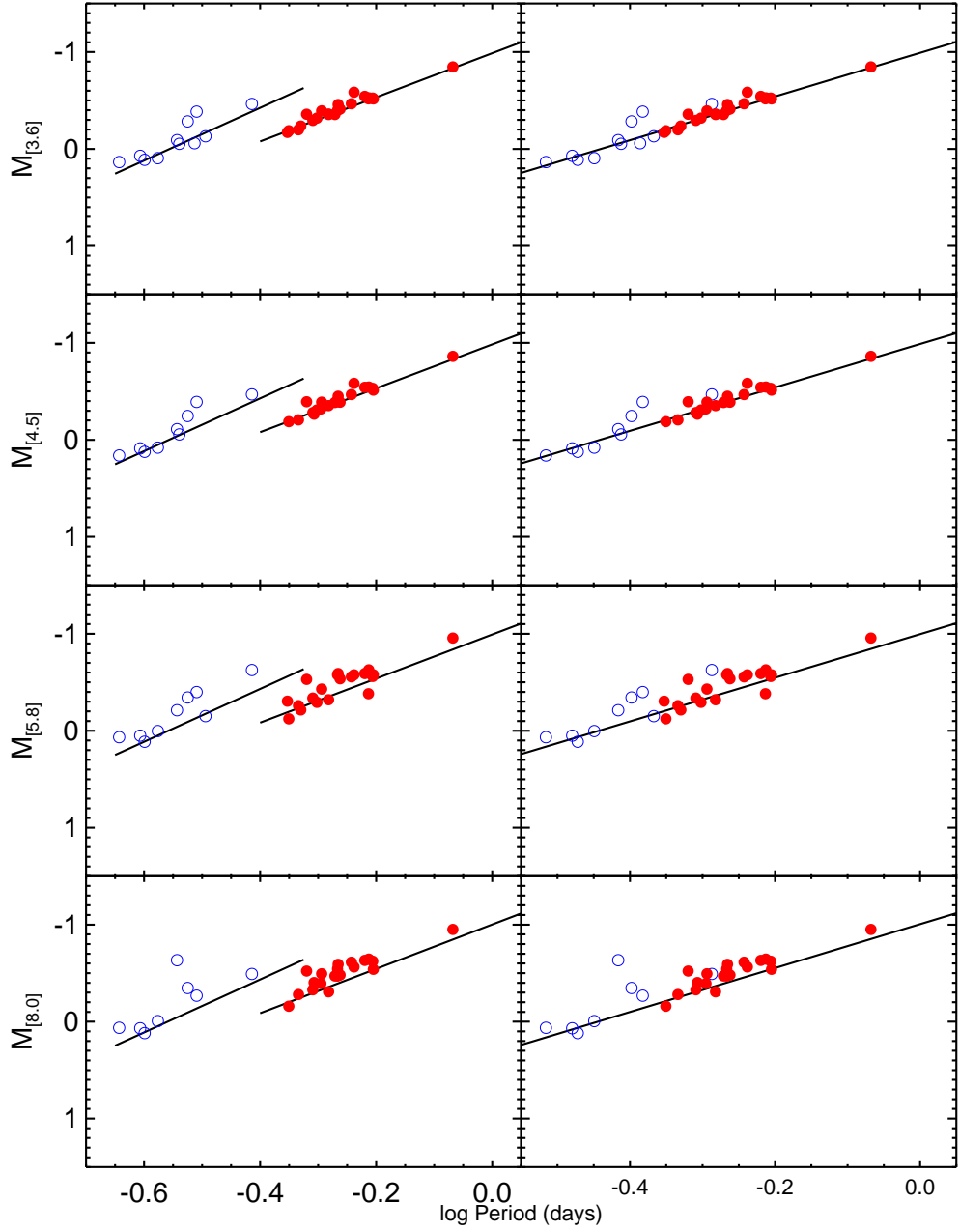}
\caption{ MIR photometry of M4 RRL corrected to absolute magnitude using best fit distance and extinction. RRc stars are shown with open blue circles and RRab stars with filled red circles. The theoretical PL relations for the cluster's metallicity is shown as the black lines. As in Figure~\ref{fig:theoryPLs}, the left panels show separate relations for the RRc and RRab stars and in the right panels the RRc stars have been fundamentalized. }
\label{fig:mirPLs}
\end{figure*}

Figure~\ref{fig:mirPLs} shows the MIR observations transformed into absolute magnitude with the best fit distance modulus and extinction. Fundamental pulsators (RRab) are shown with filled red circles while first overtone pulsators (RRc) are shown as open blue circles. In the right panels, the period of the RRc variables have been fundamentalized. The theoretical PL relation for the metallicity of the cluster ([Fe/H]=-1.1) is over-plotted. The bottom two panels show the first PL relations at the 5.8 and 8.0 $\micron$ bands. Since only single epoch observations are available for these two bands, the dispersion is larger than the anticipated intrinsic scatter of the PL relation. The data lies slightly above the theoretical line, but as expected the longer wavelengths have lower angular resolution and are much more susceptible to blending. Overall, the theoretical slope fits the data well in all four IRAC bands. 

\section{Discussion and Conclusion}\label{concl}

We have presented the first theoretical PLZ relations for RRL at MIR wavelengths. The relations were constructed from a large grid of nonlinear, time-dependent convective hydrodynamical models over a large range of metal abundances and fixed helium enrichment. We have shown that metallicity plays an important role in the zero point of these relations, and increasing the metal content decreases the zero point (i.e. RRL with higher metal abundances are fainter). With this in mind, we investigated the effect this would have on the RRL in the CRRP sample, and found that if ignored, the metallicity spread ($-2.6<$ [Fe/H] $< -0.1$) results in a scatter of 0.13 mag. When the metallicity component is accounted for, the scatter is reduced to 0.02 mag. Clearly, metallicity must be considered when analyzing a multi-metallicity sample. Consider the error budget for the Carnegie-Chicago Hubble Program, as outlined in \citet{beaton2016}. In order to obtain the projected 3\% measurement of $H_0$, it is necessary to keep the precision of distances measured via the RRL PL relation $<1.7\%$. They estimated the impact of a multiple metallicity sample using the scatter in the $H$ band PL relation in the globular cluster $\omega$ Cen. Their estimate ($\sigma_{astro}=0.06$ mag) is half the scatter we observe in the CRRP Galactic RRL sample (although we note that $\omega$ Cen covers a different range in [Fe/H] than the CRRP sample). Propagating our larger value of $\sigma_{astro}=0.13$ mag corresponds to a $3\%$ error on distances, too large for our desired uncertainty in $H_0$. To reduce the uncertainty, we must empirically calibrate the metallicity component. CRRP offers this opportunity. The globular clusters in the CRRP sample offer the necessary statistics to nail down the period dependence of the RRL PL relation, and test for any metallicity dependence of such period slope. Once the period dependence is fixed, the Galactic RRL, for which we will have milli-arcsecond parallax measurements from future \emph{Gaia} releases, will be used to calibrate the zero point and metallicity dependence. Furthermore, we will utilize multi-wavelength data in order to fit the extinction of individual objects. It is important to note that since the metallicity coefficient of the PLZ relation is wavelength independent, multi-wavelength data cannot be used to constrain the metallicity as well as the distance of individual RRL. Instead you must have a prior measurement of [Fe/H] with a maximum uncertainty of $\pm0.1$ dex in order to measure the distance of a single RRL to 2\% precision. Currently, the metallicity measurements available in the literature are very inhomogeneous. They are measured through a variety of methods and placed on different scales, and achieving our required accuracy is unlikely. Therefore, we have undertaken a program to provide homogenous measurements of [Fe/H] from high-resolution spectra for all of the RRL in our sample.  

We also present a method to fit the distance modulus and extinction of individual RRL by comparing multi-wavelength observations to our theoretical PLZ relations. At this point the error bars on the TGAS distances are too large to make any meaningful comparison with our results for the full CRRP sample. We can however directly compare our distances to those obtained by \citet{benedict2011} with \emph{HST} for five stars. We find that there is $>1\sigma$ disagreement between the theoretical and observational results for two of the five stars, UV Oct and RR Lyr. For these two stars, we see both an offset and a trend with wavelength in the residuals between observations and theory. The trend with wavelength when using the \citet{benedict2011} parallaxes suggests an incorrect value for extinction, but even when this is corrected for, the offset persists. We offer two possible explanations for this offset. The first possibility is that the parallax determined in \citet{benedict2011} is incorrect for these two stars. We find it unlikely that some unknown systematics are affecting the parallax measurement of individual stars, but note that the \citet{benedict2011} parallax measurement for UV Oct is in greater than $1 \sigma$ disagreement with the TGAS distance, and that there are typographical errors in both the parallax published for RZ Cep.\footnote{Two different values for the parallax of RZ Cep are reported in the text and Table 8 of \citet{benedict2011}. We use the value found in the text, $\pi_{abs} = 2.54 \pm 0.19$ mas, which is consistent with the distance modulus reported in their Table 8, and is more consistent with our PLZ relations.} The second possibility is that the MIR photometry for RR Lyr and UV Oct is inconsistent with the average behavior of these stars. Both of these stars are Blazhko variables and exhibit long term amplitude modulations in their optical light curves. Since the \emph{Spitzer} photometry was collected over a single pulsation phase, the average magnitude derived from one cycle may be different than one derived over many cycles. However, the \emph{WISE} photometry is randomly sampled over many different pulsation cycles, and we do not see any significant offset between the \emph{Spitzer} and \emph{WISE} mean magnitudes. Additionally, XZ Cyg is also a Blazhko star and shows no offset between the distance derived from theory and by parallax measurement. 

Our synthetic PLZ relations agree well with empirical PL and PLZ relations measured both in Galactic globular clusters and in halo RRL. These relations demonstrate the potential of RRL to be high-precision distance indicators, particularly at MIR wavelengths where the effect of extinction and intrinsic dispersion (after removing the metallicity dependence) are smallest. We also fit multi-wavelength observations of RRL in the Galactic field and in M4 to the theoretical PLZ relations to provide new distance and extinction estimates. The distance moduli of the Galactic RRL are consistent with preliminary parallax measurements from the \emph{Gaia} mission.  In M4, we fit an averaged distance modulus and extinction for the cluster, resulting in $\mu_0 = 11.257 \pm 0.035$ and $A_V = 1.45 \pm 0.12$. Both of these values are consistent with estimates from a variety of other methods, further implying the validity of the theoretical PLZ relations. 
\acknowledgments
We thank the anonymous referee for helpful comments that improved
the applicability of this manuscript.

This work has made use of data from the European Space Agency (ESA)
mission {\it Gaia} (\url{https://www.cosmos.esa.int/gaia}), processed by
the {\it Gaia} Data Processing and Analysis Consortium (DPAC,
\url{https://www.cosmos.esa.int/web/gaia/dpac/consortium}). Funding
for the DPAC has been provided by national institutions, in particular
the institutions participating in the {\it Gaia} Multilateral Agreement.

This work is based in part on archival data obtained with the Spitzer 
Space Telescope, which is operated by the Jet Propulsion Laboratory, 
California Institute of Technology under a contract with NASA. Support 
for this work was provided by an award issued by JPL/Caltech.

This publication makes use of data products from the Wide-field Infrared 
Survey Explorer, which is a joint project of the University of California, 
Los Angeles, and the Jet Propulsion Laboratory/California Institute of 
Technology, funded by the National Aeronautics and Space Administration.

%

\vspace{5mm}
\facilities{\emph{Spitzer} (IRAC), WISE, Gaia}

\clearpage

\clearpage

\scriptsize
\begin{deluxetable}{lcc cccc cccc}
\tablewidth{0pt}
\tabletypesize{\scriptsize}
\tablecaption{Intensity mean magnitudes  for entire grid of FO models \label{tab:fo-means}}
\tablehead{
\colhead{$T_e$} & \colhead{$\log L$}& \colhead{$\log P$}& \colhead{$[3.6]$}& \colhead{$[4.5]$}& \colhead{$[5.8]$}& \colhead{$[8.0]$}& \colhead{$W1$}& \colhead{$W2$}& \colhead{$W3$} & \colhead{$W4$}  \\
}
\startdata
\multicolumn{11}{c}{$Z=0.0001~Y=0.245~M=0.80 M_\sun$}  \\
7200	&	1.7600	&	-0.5088	&	-0.381	&	-0.385	&	-0.388	&	-0.393	&	-0.379	&	-0.385	&	-0.399	&	-0.415	\\
7100	&	1.7600	&	-0.4901	&	-0.426	&	-0.430	&	-0.434	&	-0.439	&	-0.424	&	-0.431	&	-0.445	&	-0.461	\\
7000	&	1.7600	&	-0.4703	&	-0.475	&	-0.479	&	-0.483	&	-0.488	&	-0.473	&	-0.479	&	-0.495	&	-0.510	\\
6900	&	1.7600	&	-0.4495	&	-0.525	&	-0.529	&	-0.533	&	-0.538	&	-0.522	&	-0.529	&	-0.544	&	-0.560	\\
6800	&	1.7600	&	-0.4282	&	-0.574	&	-0.578	&	-0.582	&	-0.587	&	-0.573	&	-0.579	&	-0.595	&	-0.611	\\
6700	&	1.7600	&	-0.4077	&	-0.622	&	-0.626	&	-0.630	&	-0.636	&	-0.621	&	-0.627	&	-0.643	&	-0.660	\\
6600	&	1.7600	&	-0.3855	&	-0.669	&	-0.673	&	-0.678	&	-0.682	&	-0.667	&	-0.674	&	-0.690	&	-0.707	\\
7100	&	1.8600	&	-0.4093	&	-0.679	&	-0.683	&	-0.687	&	-0.692	&	-0.677	&	-0.683	&	-0.698	&	-0.714	\\
7000	&	1.8600	&	-0.3891	&	-0.726	&	-0.730	&	-0.734	&	-0.739	&	-0.723	&	-0.730	&	-0.746	&	-0.762	\\
6900	&	1.8600	&	-0.3691	&	-0.774	&	-0.779	&	-0.783	&	-0.788	&	-0.772	&	-0.779	&	-0.794	&	-0.811	\\
6800	&	1.8600	&	-0.3472	&	-0.823	&	-0.828	&	-0.832	&	-0.837	&	-0.821	&	-0.828	&	-0.844	&	-0.860	\\
6700	&	1.8600	&	-0.3261	&	-0.872	&	-0.877	&	-0.881	&	-0.887	&	-0.870	&	-0.877	&	-0.893	&	-0.910	\\
\enddata
\end{deluxetable}


\scriptsize
\begin{deluxetable}{lcc cccc cccc}
\tablewidth{0pt}
\tabletypesize{\scriptsize}
\tablecaption{Intensity mean magnitudes for entire grid of FU models. \label{tab:fu-means}}
\tablehead{
\colhead{$T_e$} & \colhead{$\log L$}& \colhead{$\log P$}& \colhead{$[3.6]$}& \colhead{$[4.5]$}& \colhead{$[5.8]$}& \colhead{$[8.0]$}& \colhead{$W1$}& \colhead{$W2$}& \colhead{$W3$} & \colhead{$W4$}  \\
}
\startdata
\multicolumn{11}{c}{$Z=0.0001~Y=0.245~M=0.80 M_\sun$}  \\
6800	&	1.7600	&	-0.3016	&	-0.570	&	-0.574	&	-0.578	&	-0.584	&	-0.567	&	-0.574	&	-0.590	&	-0.607	\\
6700	&	1.7600	&	-0.2795	&	-0.624	&	-0.628	&	-0.633	&	-0.638	&	-0.621	&	-0.629	&	-0.645	&	-0.662	\\
6600	&	1.7600	&	-0.2574	&	-0.677	&	-0.682	&	-0.686	&	-0.692	&	-0.675	&	-0.682	&	-0.699	&	-0.716	\\
6500	&	1.7600	&	-0.2358	&	-0.730	&	-0.734	&	-0.739	&	-0.745	&	-0.727	&	-0.734	&	-0.751	&	-0.769	\\
6400	&	1.7600	&	-0.2124	&	-0.781	&	-0.785	&	-0.790	&	-0.796	&	-0.778	&	-0.785	&	-0.803	&	-0.821	\\
6300	&	1.7600	&	-0.1892	&	-0.829	&	-0.833	&	-0.838	&	-0.844	&	-0.826	&	-0.834	&	-0.851	&	-0.870	\\
6200	&	1.7600	&	-0.1650	&	-0.872	&	-0.877	&	-0.882	&	-0.888	&	-0.869	&	-0.877	&	-0.895	&	-0.914	\\
6100	&	1.7600	&	-0.1421	&	-0.912	&	-0.917	&	-0.921	&	-0.928	&	-0.909	&	-0.917	&	-0.935	&	-0.954	\\
6000	&	1.7600	&	-0.1178	&	-0.944	&	-0.949	&	-0.954	&	-0.960	&	-0.941	&	-0.949	&	-0.968	&	-0.987	\\
6900	&	1.8600	&	-0.2389	&	-0.765	&	-0.770	&	-0.774	&	-0.779	&	-0.763	&	-0.770	&	-0.786	&	-0.802	\\
6800	&	1.8600	&	-0.2180	&	-0.819	&	-0.824	&	-0.828	&	-0.834	&	-0.816	&	-0.824	&	-0.840	&	-0.857	\\
6600	&	1.8600	&	-0.1744	&	-0.928	&	-0.932	&	-0.936	&	-0.942	&	-0.924	&	-0.932	&	-0.949	&	-0.966	\\
6400	&	1.8600	&	-0.1282	&	-1.029	&	-1.034	&	-1.039	&	-1.045	&	-1.026	&	-1.034	&	-1.052	&	-1.070	\\
6200	&	1.8600	&	-0.0813	&	-1.122	&	-1.128	&	-1.132	&	-1.139	&	-1.119	&	-1.128	&	-1.146	&	-1.164	\\
6100	&	1.8600	&	-0.0567	&	-1.166	&	-1.172	&	-1.177	&	-1.183	&	-1.163	&	-1.172	&	-1.190	&	-1.209	\\
6000	&	1.8600	&	-0.0334	&	-1.206	&	-1.212	&	-1.217	&	-1.223	&	-1.203	&	-1.212	&	-1.231	&	-1.250	\\
5900	&	1.8600	&	-0.0088	&	-1.240	&	-1.246	&	-1.251	&	-1.258	&	-1.237	&	-1.246	&	-1.265	&	-1.284	\\
\enddata
\end{deluxetable}


\scriptsize
\begin{deluxetable}{l rrrc rrrc rrrc}
\tablewidth{0pt}
\tabletypesize{\scriptsize}
\tablecaption{Theoretical NIR and MIR Period--Luminosity relations for RR Lyrae. \label{tab:PLZ}}
\tablehead{
\colhead{Filters\tablenotemark{a}} & \colhead{a}& \colhead{b}& \colhead{c}& \colhead{$\sigma$}& \colhead{a}& \colhead{b}& \colhead{c}& \colhead{$\sigma$}& \colhead{a} & \colhead{b} & \colhead{c} & \colhead{$\sigma$} \\
                    & mag & mag & mag & mag & mag  & mag  & mag & mag & mag  & mag & mag & mag
}
\startdata
          & \multicolumn{4}{c}{FO\tablenotemark{b}} & \multicolumn{4}{c}{FU\tablenotemark{b}} & \multicolumn{4}{c}{FU$+$FO \tablenotemark{c}} \\
\multicolumn{13}{c}{--- Spitzer ---} \\
I1  & -1.344 & -2.718 &  0.152 & 0.021 & -0.786 & -2.276 &  0.184 & 0.035 & -0.793 & -2.251 &  0.180 & 0.037 \\
       & $\pm$0.024 & $\pm$0.046 & $\pm$0.004 & & $\pm$0.007 & $\pm$0.021 & $\pm$0.004 & & $\pm$0.007 & $\pm$0.018 & $\pm$0.003 \\
I2  & -1.348 & -2.720 &  0.153 & 0.021 & -0.775 & -2.262 &  0.190 & 0.036 & -0.785 & -2.239 &  0.185 & 0.038 \\
       & $\pm$0.024 & $\pm$0.046 & $\pm$0.004 & & $\pm$0.007 & $\pm$0.022 & $\pm$0.004 & & $\pm$0.007 & $\pm$0.018 & $\pm$0.003 \\
I3  & -1.352 & -2.724 &  0.153 & 0.021 & -0.786 & -2.273 &  0.188 & 0.035 & -0.795 & -2.250 &  0.184 & 0.037 \\
       & $\pm$0.023 & $\pm$0.046 & $\pm$0.004 & & $\pm$0.007 & $\pm$0.021 & $\pm$0.004 & & $\pm$0.007 & $\pm$0.018 & $\pm$0.003 \\
I4  & -1.355 & -2.728 &  0.155 & 0.021 & -0.798 & -2.288 &  0.186 & 0.035 & -0.805 & -2.264 &  0.183 & 0.036 \\
       & $\pm$0.023 & $\pm$0.046 & $\pm$0.004 & & $\pm$0.007 & $\pm$0.021 & $\pm$0.004 & & $\pm$0.007 & $\pm$0.017 & $\pm$0.003 \\
\multicolumn{13}{c}{--- WISE ---} \\
W1  & -1.341 & -2.716 &  0.152 & 0.021 & -0.784 & -2.274 &  0.183 & 0.036 & -0.790 & -2.247 &  0.180 & 0.037 \\
       & $\pm$0.024 & $\pm$0.047 & $\pm$0.004 & & $\pm$0.007 & $\pm$0.022 & $\pm$0.004 & & $\pm$0.007 & $\pm$0.018 & $\pm$0.003 \\
W2  & -1.348 & -2.720 &  0.153 & 0.021 & -0.774 & -2.261 &  0.190 & 0.036 & -0.784 & -2.237 &  0.185 & 0.038 \\
       & $\pm$0.024 & $\pm$0.046 & $\pm$0.004 & & $\pm$0.008 & $\pm$0.022 & $\pm$0.004 & & $\pm$0.007 & $\pm$0.018 & $\pm$0.003 \\
W3  & -1.357 & -2.731 &  0.157 & 0.021 & -0.800 & -2.292 &  0.188 & 0.035 & -0.807 & -2.267 &  0.185 & 0.036 \\
       & $\pm$0.023 & $\pm$0.045 & $\pm$0.004 & & $\pm$0.007 & $\pm$0.021 & $\pm$0.004 & & $\pm$0.007 & $\pm$0.018 & $\pm$0.003 \\
W4  & -1.355 & -2.735 &  0.166 & 0.020 & -0.799 & -2.298 &  0.196 & 0.034 & -0.805 & -2.274 &  0.193 & 0.036 \\
       & $\pm$0.022 & $\pm$0.044 & $\pm$0.004 & & $\pm$0.007 & $\pm$0.021 & $\pm$0.004 & & $\pm$0.007 & $\pm$0.017 & $\pm$0.003 \\
\enddata
\tablenotetext{a}{I1, I2, I3, and I4 correspond to IRAC 3.6 $\micron$, 4.5 $\micron$, 5.8 $\micron$, and 8.0 $\micron$, respectively}
\tablenotetext{b}{The PLZ relations are of the form: 
$M_X$= a + b$\times$$\log\ P$ + c$\times$$[Fe/H]$.}
\tablenotetext{c}{The periods of FO variables were fundamentalized with the relation: $\log P_F$=$\log P_{FO}$+0.127. }
\end{deluxetable}


\scriptsize
\begin{deluxetable}{l c cccc cccc cccc}
\tablewidth{0pt}
\tabletypesize{\scriptsize}
\tablecaption{Theoretical 2-band Period-Wesenheit-Metallicity relations \label{tab:PWZ2} }
\tablehead{
\colhead{Filters\tablenotemark{a}} & \colhead{$\alpha$} & \colhead{a} & \colhead{b} & \colhead{c} & \colhead{$\sigma$} & \colhead{a} & \colhead{b} & \colhead{c} & \colhead{$\sigma$}
& \colhead{a} & \colhead{b} & \colhead{c} & \colhead{$\sigma$}
}
\startdata
      &    & \multicolumn{4}{c}{FO} & \multicolumn{4}{c}{FU} & \multicolumn{4}{c}{FU$+$FO} \\
 I1,I-I1  & 0.126 & -1.437 & -2.784 &  0.154 & 0.018 & -0.882 & -2.354 &  0.186 & 0.030 & -0.884 & -2.343 &  0.182 & 0.032 \\
         &   & $\pm$0.020 & $\pm$0.039 & $\pm$0.003 & & $\pm$0.006 & $\pm$0.018 & $\pm$0.003 & & $\pm$0.006 & $\pm$0.015 & $\pm$0.003 \\
 I2,I-I2  & 0.094 & -1.419 & -2.770 &  0.154 & 0.018 & -0.848 & -2.319 &  0.193 & 0.032 & -0.852 & -2.306 &  0.187 & 0.034 \\
         &   & $\pm$0.021 & $\pm$0.041 & $\pm$0.003 & & $\pm$0.007 & $\pm$0.019 & $\pm$0.003 & & $\pm$0.006 & $\pm$0.016 & $\pm$0.003 \\
\enddata
\tablenotetext{a}{I1, I2, I3, and I4 correspond to IRAC 3.6 $\micron$, 4.5 $\micron$, 5.8 $\micron$, and 8.0 $\micron$, respectively}
\end{deluxetable}


\scriptsize
\begin{deluxetable}{l c cccc cccc cccc}
\tablewidth{0pt}
\tabletypesize{\scriptsize}
\tablecaption{Theoretical 3-band Period-Wesenheit-Metallicity relations \label{tab:PWZ3} }
\tablehead{
\colhead{Filters\tablenotemark{a}} & \colhead{$\alpha$} & \colhead{a} & \colhead{b} & \colhead{c} & \colhead{$\sigma$} & \colhead{a} & \colhead{b} & \colhead{c} & \colhead{$\sigma$}
& \colhead{a} & \colhead{b} & \colhead{c} & \colhead{$\sigma$}
}
\startdata
      &    & \multicolumn{4}{c}{FO} & \multicolumn{4}{c}{FU} & \multicolumn{4}{c}{FU$+$FO} \\
   V,B-I1  & 0.793 & -1.448 & -2.750 &  0.119 & 0.023 & -0.943 & -2.439 &  0.131 & 0.033 & -0.935 & -2.382 &  0.134 & 0.036 \\
         &   & $\pm$0.026 & $\pm$0.051 & $\pm$0.004 & & $\pm$0.007 & $\pm$0.020 & $\pm$0.004 & & $\pm$0.007 & $\pm$0.017 & $\pm$0.003 \\
  V,B-I2  & 0.783 & -1.435 & -2.741 &  0.120 & 0.023 & -0.917 & -2.413 &  0.137 & 0.032 & -0.910 & -2.355 &  0.138 & 0.035 \\
         &   & $\pm$0.026 & $\pm$0.051 & $\pm$0.004 & & $\pm$0.007 & $\pm$0.019 & $\pm$0.003 & & $\pm$0.007 & $\pm$0.017 & $\pm$0.003 \\
\enddata
\tablenotetext{a}{I1, I2, I3, and I4 correspond to IRAC 3.6 $\micron$, 4.5 $\micron$, 5.8 $\micron$, and 8.0 $\micron$, respectively}
\end{deluxetable}

\scriptsize
\begin{deluxetable}{l cccc r ccccc }
\tablewidth{0pt}
\tabletypesize{\scriptsize}
\tablecaption{Galactic RR Lyrae sample \label{tab:stars}}
\tablehead{
\colhead{Name} & \colhead{RA} & \colhead{Dec} & \colhead{Period} & \colhead{[Fe/H]\tablenotemark{a}} & \colhead{Type} & \colhead{$A_V$\tablenotemark{a}} & \colhead{$HST~\mu$} & \colhead{$Gaia~\mu$} & \colhead{$\mu$} & \colhead{$A_V$}
}
\startdata
AB	UMa	&   12:11:14.58669	&	47:49:43.8088	&  0.59958113	&	-0.49	&	RRab	&  $ 	0.07	$ &				   &  $	10.146	\pm	0.617	$ &  $ 	 9.984  \pm	0.020	$ & $  0.405	\pm	0.108 	$ \\
AE	Boo	&   14:47:35.26451	&	16:50:43.5538	&  0.31489000	&	-1.39	&	RRc		&  $ 	0.07	$ &				   &  $	9.584	\pm	0.473	$ &  $ 	 9.920  \pm	0.032	$ & $  0.216	\pm	0.168 	$ \\
AM	Tuc	&   01:18:30.64730	&	-67:55:04.9522	&  0.40580160	&	-1.49	&	RRc		&  $ 	0.07	$ &				   &  $	10.353	\pm	0.661	$ &  $ 	11.036 \pm	0.024	$ & $  0.162	\pm	0.158 	$ \\
AN	Ser	&   15:53:31.05070	&	12:57:40.1293	&  0.52207144	&	-0.07	&	RRab	&  $ 	0.12	$ &				   &  $	10.565	\pm	0.802	$ &  $ 	 9.950  \pm	0.024	$ & $  0.328	\pm	0.158 	$ \\
AP	Ser	&   15:14:00.92200	&	09:58:51.8100	&  0.34083000	&	-1.58	&	RRc		&  $ 	0.13	$ &				   & 	\nodata			  		   &  $	10.500 \pm	0.024	$ & $  0.251	\pm	0.159 	$ \\
AV	Peg	&   21:52:02.79453	&	22:34:29.3787	&  0.39037470	&	-0.08	&	RRc		&  $ 	0.21	$ &				   &  $	9.084	\pm	0.326	$ &  $ 	 9.186  \pm	0.021	$ & $  0.419	\pm	0.111 	$ \\
BB	Pup	&   08:24:22.65000	&	-19:32:31.4000	&  0.48054884	&	-0.60	&	RRab	&   \nodata   &				   & 	\nodata			  		   &  $	11.027 \pm	0.029   	$ & $  0.473	\pm	0.119 	$ \\
BH	Peg	&   22:53:01.03678	&	15:47:16.5999	&  0.64099300	&	-1.22	&	RRab	&  $ 	0.24	$ &				   &  $	9.272	\pm	0.343	$ &  $ 	 9.540  \pm	0.023	$ & $  0.544	\pm	0.156 	$ \\
BX	Leo	&   11:38:02.06518	&	16:32:36.1864	&  0.36275500	&	-1.28	&	RRc		&  $ 	0.07	$ &				   &  $	11.366	\pm	1.124	$ &  $ 	11.023 \pm	0.021	$ & $  0.076	\pm	0.109 	$ \\
CS	Eri	&   02:37:05.75874	&	-42:57:48.0588	&  0.31133100	&	-1.41	&	RRc		&  $ 	0.06	$ &				   &  $	8.331	\pm	0.229	$ &  $ 	 8.292  \pm	0.022	$ & $  <0.153* 	                		$ \\
CU	Com	&   12:24:46.61000	&	22:24:28.2000	&  0.40576050	&	-2.38	&	RRc		&  \nodata	  &				   & 	\nodata			  		   &  $	12.570 \pm	0.026	$ & $  0.313	\pm	0.179 	$ \\
DH	Peg	&   22:15:25.64000	&	06:49:21.4500	&  0.25551053	&	-1.24	&	RRc		&  \nodata	  &				   & 	\nodata			  		   &  $	 8.528  \pm	0.022	$ & $  0.351	\pm	0.152 	$ \\
DX	Del	&   20:47:28.35586	&	12:27:50.6919	&  0.47261673	&	-0.39	&	RRab	&  $ 	0.29	$ &				   &  $	8.898	\pm	0.293	$ &  $ 	 8.756  \pm	0.022	$ & $  0.622	\pm	0.148 	$ \\
HK	Pup	&   07:44:46.80188	&	-13:05:56.3068	&  0.73420730	&	-1.11	&	RRab	&  $ 	0.50	$ &				   &  $	11.387	\pm	1.077	$ &  $ 	10.536 \pm	0.024	$ & $  0.472	\pm	0.158 	$ \\
MT	Tel	&   19:02:12.28034	&	-46:39:12.0870	&  0.31689740	&	-1.85	&	RRc		&  $ 	0.12	$ &				   &  $	9.228	\pm	0.471	$ &  $ 	 8.356  \pm	0.022	$ & $  0.206	\pm	0.153 	$ \\
RR	Cet	&   01:32:08.17309	&	01:20:30.2342	&  0.55302900	&	-1.45	&	RRab	&  $ 	0.07	$ &				   & 	\nodata			  		   &  $	 8.966  \pm	0.020	$ & $  0.308	\pm	0.109 	$ \\
RR	Gem	&   07:21:33.53246	&	30:52:59.4617	&  0.39729000	&	-0.29	&	RRab	&  $ 	0.17	$ &				   &  $	10.648	\pm	0.757	$ &  $ 	10.146 \pm	0.021	$ & $  0.386	\pm	0.111    	$ \\
RR	Leo	&   10:07:43.46008	&	23:59:30.3292	&  0.45239330	&	-1.60	&	RRab	&  $ 	0.11	$ &				   & 	\nodata			  		   &  $	 9.942  \pm	0.021	$ & $  0.303	\pm	0.110 	$ \\
RR	Lyr	&   19:25:27.91285	&	42:47:03.6942	&  0.56683780	&	-1.39	&	RRab	&  $ 	0.09	$ &  $ 7.14 \pm 0.07 $ &  $ 	7.195	\pm	0.139	$ &  $ 	 6.941  \pm	0.019   	$ & $  0.392	\pm	0.106 	$ \\
RU	Psc	&   01:14:26.03793	&	24:24:56.3725	&  0.39036500	&	-1.75	&	RRc		&  $ 	0.13	$ &				   &  $	9.216	\pm	0.446	$ &  $ 	 9.553  \pm	0.021	$ & $  0.160	\pm	0.111 	$ \\
RU	Scl	&   00:02:48.10957	&	-24:56:43.0689	&  0.49335500	&	-1.27	&	RRab	&  $ 	0.06	$ &				   &  $	9.088	\pm	0.423	$ &  $ 	 9.477  \pm	0.023	$ & $  0.372	\pm	0.156 	$ \\
RV	CrB	&   16:19:25.85137	&	29:42:47.6404	&  0.33168000	&	-1.69	&	RRc		&  $ 	0.12	$ &				   &  $	11.514	\pm	1.229	$ &  $ 	10.772 \pm	0.024	$ & $  <0.159* 	                		$ \\
RV	Oct	&   13:46:31.74979	&	-84:24:06.3861	&  0.57116250	&	-1.71	&	RRab	&  $ 	0.56	$ &				   &  $	9.725	\pm	0.444	$ &  $ 	 9.991  \pm	0.027	$ & $  0.604	\pm	0.117 	$ \\
RV	UMa	&   13:33:18.08438	&	53:59:14.6081	&  0.46806000	&	-1.20	&	RRab	&  $ 	0.06	$ &				   &  $	10.651	\pm	0.676	$ &  $ 	10.037  \pm	0.021	$ & $  0.235	\pm	0.109 	$ \\
RX	Eri	&   04:49:44.29135	&	-15:44:28.2502	&  0.58724622	&	-1.33	&	RRab	&  $ 	0.18	$ &				   &  $	8.69		\pm	0.337	$ &  $ 	 8.822  \pm	0.020	$ & $  0.451	\pm	0.111 	$ \\
RZ	Cep	&   22:39:13.17772	&	64:51:30.6036	&  0.30868000	&	-1.77	&	RRc		&  $ 	0.24	$ &  $ 8.03 \pm 0.16 $ &  $ 	7.88		\pm	0.193	$ &  $ 	 8.060  \pm	0.020	$ & $  0.739	\pm	0.110 	$ \\
ST	Boo	&   15:30:39.23085	&	35:47:04.3057	&  0.62228600	&	-1.76	&	RRab	&  $ 	0.07	$ &				   &  $	10.809	\pm	1.027	$ &  $ 	10.507 \pm	0.021	$ & $  0.165	\pm	0.109 	$ \\
ST	CVn	&   13:57:34.06087	&	29:51:28.6887	&  0.32904500	&	-1.07	&	RRc		&  $ 	0.04	$ &				   &  $	13.249	\pm	3.478	$ &  $ 	10.607 \pm	0.024	$ & $  0.115	\pm	0.159 	$ \\
SU	Dra	&   11:37:56.60743	&	67:19:47.0633	&  0.66042001	&	-1.80	&	RRab	&  $ 	0.03	$ &  $ 9.35 \pm 0.24 $ &  $       	9.223	\pm	0.431	$ & $ 	 9.301  \pm	0.020	$ & $  0.195  \pm	0.109 	$ \\
SV	Eri	&   03:11:52.10655	&	-11:21:14.0708	&  0.71385300	&	-1.70	&	RRab	&  $ 	0.26	$ &				   &  $	9.371	\pm	0.427	$ & $ 	 9.308  \pm	0.022	$ & $  0.478  \pm	0.154 	$ \\
SV	Hya	&   12:30:30.50357	&	-26:02:51.1231	&  0.47854280	&	-1.50	&	RRab	&  $ 	0.25	$ &				   &  $	9.397	\pm	0.401	$ & $ 	 9.683  \pm	0.023	$ & $  0.410  \pm	0.157 	$ \\
SV	Scl	&   01:44:59.66250	&	-30:03:33.3859	&  0.37735039	&	-1.77	&	RRc		&  $ 	0.04	$ &				   &  $	11.783	\pm	1.593	$ & $ 	10.950 \pm	0.024	$ & $  0.073  \pm	0.158 	$ \\
SW	And	&   00:23:43.08936	&	29:24:03.6365	&  0.44226020	&	-0.24	&	RRab	&  $ 	0.12	$ &				   &  $	8.766	\pm	0.321	$ & $ 	 8.504  \pm	0.020	$ & $  0.409  \pm	0.108 	$ \\
SW	Dra	&   12:17:46.63152	&	69:30:38.2236	&  0.56966993	&	-1.12	&	RRab	&  $ 	0.04	$ &				   &  $	9.978	\pm	0.496	$ & $ 	 9.726  \pm	0.021	$ & $  0.241  \pm	0.112 	$ \\
SX	UMa	&   13:26:13.46025	&	56:15:25.0606	&  0.30711780	&	-1.81	&	RRc		&  $ 	0.03	$ &				   &  $	11.91	\pm	1.804	$ & $ 	10.328 \pm	0.022	$ & $  0.001  \pm	0.109    	$ \\
T	Sex	&   09:53:28.39930	&	02:03:26.3563	&  0.32468460	&	-1.34	&	RRc		&  $ 	0.14	$ &				   & 	\nodata			  		   & $		 9.358  \pm	0.021	$ & $  0.128  \pm	0.110 	$ \\
TT	Lyn	&   09:03:07.78856	&	44:35:08.1213	&  0.59743436	&	-1.56	&	RRab	&  $ 	0.05	$ &				   &  $	9.078	\pm	0.569	$ & $ 	 9.138  \pm	0.020	$ & $  0.316  \pm	0.109 	$ \\
TU	UMa	&   11:29:48.49055	&	30:04:02.4094	&  0.55765870	&	-1.51	&	RRab	&  $ 	0.07	$ &				   &  $	9.166	\pm	0.463	$ & $ 	 9.101  \pm	0.020	$ & $  0.284  \pm	0.108 	$ \\
TV	Boo	&   14:16:36.58091	&	42:21:35.6927	&  0.31256107	&	-2.44	&	RRc		&  $ 	0.03	$ &				   &  $	10.607	\pm	0.669	$ & $ 	10.580 \pm	0.021	$ & $  <0.111*                  		$ \\
TW	Her	&   17:54:31.19965	&	30:24:37.7117	&  0.39960010	&	-0.69	&	RRab	&  $ 	0.13	$ &				   &  $	10.939	\pm	1.03		$ & $ 	10.224 \pm	0.024	$ & $  0.424  \pm	0.158 	$ \\
UU	Vir	&   12:08:35.07300	&	-00:27:24.3000	&  0.47560890	&	-0.87	&	RRab	&  $ 	0.06	$ &				   &  $	9.345	\pm	0.507	$ & $ 	 9.688  \pm	0.021	$ & $  0.214  \pm	0.110 	$ \\
UV	Oct	&   16:32:25.53387	&	-83:54:10.5183	&  0.54258000	&	-1.74	&	RRab	&  $ 	0.28	$ &  $ 8.87 \pm 0.13 $ &  $ 	8.478	\pm	0.242	$ & $ 	 8.660  \pm	0.022	$ & $  0.603  \pm	0.153 	$ \\
UY	Boo	&   13:58:46.33747	&	12:57:06.4558	&  0.65083000	&	-2.56	&	RRab	&  $ 	0.10	$ &				   &  $	13.338	\pm	2.624	$ & $ 	10.556 \pm	0.020	$ & $  0.233  \pm	0.109 	$ \\
UY	Cam	&   07:58:58.88054	&	72:47:15.4203	&  0.26702740	&	-1.33	&	RRc		&  $ 	0.07	$ &				   &  $ 	9.672	\pm	1.203	$ & $ 	10.804 \pm	0.024	$ & $  0.134  \pm	0.160 	$ \\
UY	Cyg	&   20:56:28.30246	&	30:25:40.3104	&  0.56070478	&	-0.80	&	RRab	&  $ 	0.40	$ &				   & 	\nodata			  		   & $		10.078 \pm	0.021	$ & $  0.471  \pm	0.110 	$ \\
V	Ind	&   21:11:29.90402	&	-45:04:28.3835	&  0.47960170	&	-1.50	&	RRab	&  $ 	0.13	$ &				   &  $	9.785	\pm	0.537	$ & $ 	 9.179  \pm	0.023	$ & $  0.492  \pm	0.155 	$ \\
V440  Sgr	&   19:32:20.78211	&	-23:51:12.7553	&  0.47747883	&	-1.40	&	RRab	&  $ 	0.26	$ &				   &  $	9.467	\pm	0.708	$ & $ 	 9.336  \pm	0.021	$ & $  0.498  \pm	0.112 	$ \\
V675  Sgr	&   18:13:35.41006	&	-34:19:01.8403	&  0.64228930	&	-2.28	&	RRab	&  $ 	0.40	$ &				   & 	\nodata			  		   & $	 	9.691  \pm	0.023	$ & $  0.424  \pm	0.156 	$ \\
VX	Her	&   16:30:40.79990	&	18:22:00.5524	&  0.45535984	&	-1.58	&	RRab	&  $ 	0.14	$ &				   &  $	10.219	\pm	0.581	$ & $ 	 9.878  \pm	0.024	$ & $  0.331  \pm	0.158 	$ \\
W	Crt	&   11:26:29.64190	&	-17:54:51.6812	&  0.41201459	&	-0.54	&	RRab	&  $ 	0.12	$ &				   &  $	10.207	\pm	0.664	$ & $ 	10.815 \pm	0.028   	$ & $  0.168  \pm	0.119 	$ \\
WY	Ant	&   10:16:04.97700	&	-29:43:42.9000	&  0.57434560	&	-1.48	&	RRab	&  $ 	0.18	$ &				   &  $	10.41	\pm	0.598	$ & $ 	10.067  \pm	0.021	$ & $  0.402  \pm	0.110	$ \\
X	Ari	&   03:08:30.88520	&	10:26:45.2282	&  0.65117230	&	-2.43	&	RRab	&  $ 	0.56	$ &				   &  $	8.468	\pm	0.239	$ & $ 	 8.660  \pm	0.020	$ & $  0.704  \pm	0.108	$ \\
XX	And	&   01:17:27.41498	&	38:57:02.0359	&  0.72275700	&	-1.94	&	RRab	&  $ 	0.12	$ &				   & 	\nodata			  		   & $ 	10.253  \pm	0.020	$ & $  0.308  \pm	0.109	$ \\
XZ	Cyg	&   19:32:29.30486	&	56:23:17.4900	&  0.46659934	&	-1.44	&	RRab	&  $ 	0.30	$ & $	 8.98 \pm 0.22 $  &  $ 	9.03		\pm	0.314	$ & $ 	 8.946  \pm	0.021	$ & $  0.240  \pm	0.111	$ \\
YZ	Cap	&   21:19:32.41125	&	-15:07:01.1574	&  0.27345630	&	-1.06	&	RRc		&  $ 	0.20	$ &				   &  $	11.358	\pm	1.188	$ & $ 	10.325  \pm	0.022	$ & $  0.296  \pm	0.113	$ \\
\enddata
\tablenotetext{a}{Adopted from \cite{feast2008}; $A_V=3.1E(B-V)$}
\tablenotetext{*}{Fit only provided upper limit for these stars.}
\end{deluxetable}


\scriptsize
\begin{deluxetable}{l ccccc}
\tablewidth{0pt}
\tabletypesize{\scriptsize}
\tablecaption{Updated photometry for RR Lyrae in M4 \label{tab:M4data}}
\tablehead{
\colhead{Name}  & \colhead{Period} & \colhead{[3.6]} & \colhead{[4.5]} & \colhead{[5.8]} & \colhead{[8.0]}
}
\startdata
V01	&	0.28888261	& $	11.304	\pm	0.023	$ & $	11.278	\pm	0.073	$ & \nodata & \nodata \\
V02	&	0.5356819	& $	11.002	\pm	0.093	$ & $	10.949	\pm	0.031	$ & \nodata & $	10.863	\pm	0.139	$ \\
V03	&	0.50667787	& 	\nodata	 & $	11.015	\pm	0.026	$ & \nodata & $	10.941	\pm	0.135	$ \\
V05	&	0.62240112	& $	10.833	\pm	0.036	$ & $	10.803	\pm	0.029	$ & $	10.774	\pm	0.125	$ & $	10.711	\pm	0.077	$ \\
V06	&	0.3205151	& $	11.226	\pm	0.044	$ & \nodata & $	11.183	\pm	0.053	$ & \nodata\\
V07	&	0.49878722	& $	11.039	\pm	0.043	$ & $	11.026	\pm	0.032	$ & $	11.041	\pm	0.156	$ & \nodata \\
V08	&	0.50822359	& $	10.964	\pm	0.045	$ & $	10.945	\pm	0.027	$ & $	10.904	\pm	0.226	$ & $	10.84	\pm	0.141	$ \\
V09	&	0.57189447	& $	10.891	\pm	0.043	$ & $	10.867	\pm	0.039	$ & $	10.778	\pm	0.142	$ & $	10.721	\pm	0.142	$ \\
V10	&	0.49071753	& $	11.063	\pm	0.048	$ & $	11.053	\pm	0.032	$ & $	10.997	\pm	0.16	$ & $	11.005	\pm	0.16	$ \\
V11	&	0.49320868	& \nodata & $	11.069	\pm	0.03	$ & \nodata & $	10.931	\pm	0.13	$ \\
V12	&	0.4461098	& $	11.169	\pm	0.064	$ & $	11.146	\pm	0.034	$ & $	11.211	\pm	0.171	$ & $	11.175	\pm	0.171	$ \\
V14	&	0.46353111	& $	11.159	\pm	0.061	$ & $	11.128	\pm	0.039	$ & $	11.075	\pm	0.169	$ & $	11.054	\pm	0.169	$ \\
V15	&	0.44366077	& $	11.186	\pm	0.044	$ & \nodata & $	11.029	\pm	0.158	$ & \nodata \\
V16	&	0.54254824	& $	10.899	\pm	0.036	$ & $	10.884	\pm	0.027	$ & $	10.744	\pm	0.122	$ & $	10.743	\pm	0.122	$ \\
V18	&	0.47879201	& $	10.999	\pm	0.051	$ & $	10.941	\pm	0.05	$ & $	10.804	\pm	0.13	$ & $	10.812	\pm	0.13	$ \\
V19	&	0.46781108	& $	11.121	\pm	0.045	$ & \nodata & $	11.119	\pm	0.136	$ & \nodata \\
V20	&	0.30941948	& $	10.972	\pm	0.105	$ & $	10.943	\pm	0.106	$ & $	10.935	\pm	0.077	$ & $	11.066	\pm	0.039	$ \\
V21	&	0.47200742	& $	10.772	\pm	0.056	$ & $	10.771	\pm	0.05	$ & $	10.569	\pm	0.123	$ & $	10.556	\pm	0.123	$ \\
V22	&	0.60306358	& $	10.815	\pm	0.047	$ & $	10.792	\pm	0.028	$ & $	10.744	\pm	0.092	$ & $	10.702	\pm	0.092	$ \\
V23	&	0.29861557	& $	11.073	\pm	0.034	$ & $	11.088	\pm	0.021	$ & $	10.991	\pm	0.045	$ & $	10.987	\pm	0.053	$ \\
V24	&	0.54678333	& $	10.946	\pm	0.042	$ & $	10.946	\pm	0.027	$ & $	10.798	\pm	0.11	$ & $	10.853	\pm	0.11	$ \\
V25	&	0.61273479	& $	10.83	\pm	0.046	$ & $	10.789	\pm	0.031	$ & $	10.706	\pm	0.139	$ & $	10.689	\pm	0.139	$ \\
V26	&	0.54121739	& $	10.959	\pm	0.045	$ & $	10.928	\pm	0.037	$ & $	10.757	\pm	0.165	$ & $	10.79	\pm	0.165	$ \\
V27	&	0.61201829	& $	10.84	\pm	0.041	$ & \nodata & $	10.952	\pm	0.121	$ & \nodata \\
V28	&	0.52234107	& $	10.997	\pm	0.04	$ & $	10.981	\pm	0.038	$ & $	11.014	\pm	0.178	$ & $	11.026	\pm	0.178	$ \\
V29	&	0.52248466	& $	11.002	\pm	0.044	$ & \nodata & \nodata & \nodata \\
V30	&	0.26974906	& \nodata & \nodata & \nodata & \nodata \\
V31	&	0.50520423	& \nodata & \nodata & \nodata & \nodata \\
V32	&	0.57910475	& \nodata & \nodata& \nodata & \nodata \\
V33	&	0.61483542	& \nodata & \nodata & \nodata & \nodata \\
V34	&	0.5548	& \nodata & \nodata & \nodata & \nodata \\
V35	&	0.62702374	& \nodata& \nodata & \nodata & \nodata \\
V36	&	0.54130918	& \nodata & $	10.939	\pm	0.038	$ & \nodata & $	10.864	\pm	0.142	$ \\
V37	&	0.24734353	& $	11.428	\pm	0.043	$ & $	11.422	\pm	0.02	$ & $	11.384	\pm	0.026	$ & $	11.402	\pm	0.03	$ \\
V38	&	0.57784635	& $	10.772	\pm	0.047	$ & $	10.751	\pm	0.033	$ & $	10.757	\pm	0.104	$ & $	10.77	\pm	0.104	$ \\
V39	&	0.623954	& $	10.841	\pm	0.041	$ & $	10.821	\pm	0.025	$ & $	10.757	\pm	0.091	$ & $	10.794	\pm	0.091	$ \\
V40	&	0.38533005	& $	10.893	\pm	0.051	$ & $	10.865	\pm	0.034	$ & $	10.708	\pm	0.153	$ & $	10.842	\pm	0.366	$ \\
V41	&	0.2517418	& $	11.469	\pm	0.036	$ & $	11.457	\pm	0.019	$ & $	11.447	\pm	0.038	$ & $	11.452	\pm	0.038	$ \\
V42	&	0.3068549	& $	11.298	\pm	0.11	$ & \nodata & \nodata & \nodata \\
V49	&	0.22754331	& $	11.492	\pm	0.035	$ & $	11.495	\pm	0.029	$ & $	11.4	\pm	0.029	$ & $	11.397	\pm	0.029	$ \\
V52	&	0.85549784	& $	10.511	\pm	0.041	$ & $	10.473	\pm	0.034	$ & $	10.378	\pm	0.071	$ & $	10.383	\pm	0.01	$ \\
V61	&	0.26528645	& $	11.452	\pm	0.053	$ & $	11.414	\pm	0.018	$ & $	11.337	\pm	0.03	$ & $	11.328	\pm	0.031	$ \\
C01	&	0.2862573	& $	11.266	\pm	0.102	$ & $	11.224	\pm	0.11	$ & $	11.122	\pm	0.112	$ & $	10.7	\pm	0.235	$ \\
\enddata

\end{deluxetable}


\scriptsize
\begin{deluxetable}{l ccc ccc ccc}
\tablewidth{0pt}
\tabletypesize{\scriptsize}
\tablecaption{Calibrated empirical PL relations for M4 \label{tab:M4PLs}}
\tablehead{
\colhead{Band}  & \colhead{$a$\tablenotemark{a}} & \colhead{$b$\tablenotemark{a}} & \colhead{$\sigma$\tablenotemark{b}} & \colhead{$a$\tablenotemark{a}} & \colhead{$b$\tablenotemark{a}} & \colhead{$\sigma$\tablenotemark{b}} & \colhead{$a$\tablenotemark{a}} & \colhead{$b$\tablenotemark{a}} & \colhead{$\sigma$\tablenotemark{b}} 
}
\startdata
          & \multicolumn{3}{c}{FO} & \multicolumn{3}{c}{FU} & \multicolumn{3}{c}{FU$+$FO} \\
$[3.6]$ & $-1.008\pm0.170$ &  $-2.75 \pm 0.42$ & 0.075 & $-1.155\pm0.089$ & $-2.34\pm0.14$ & 0.040 & $-1.112 \pm 0.089$ & $-2.30 \pm  0.11$ & 0.055 \\
$[4.5]$ & $-1.032\pm0.170$ &  $-3.00\pm0.33$ & 0.056 & $-1.170\pm0.089$ & $-2.36\pm0.17$ &  0.044 & $-1.139 \pm 0.089$ & $-2.34 \pm 0.10$ & 0.053\\
$[5.8]$ & $-1.220\pm$0.29  &  $-3.18\pm0.56$ & 0.10  & $-1.239\pm0.11 $    & $-2.43\pm0.34$ &  0.096 & $-1.190\pm0.09 $& $-2.34 \pm  0.20$ &  0.10\\
$[8.0]$ & $-1.263\pm$0.55 &  $-3.13\pm1.20$ &  0.20 &  $-1.270\pm0.10 $   & $-2.45\pm0.28$ &  0.074 & $-1.187\pm0.10 $  & $-2.22 \pm 0.25$ & 0.12\\
\enddata
\tablenotetext{a}{$m=a+b \log P$}
\tablenotetext{b}{Dispersion of RRL in the cluster M4.}
\end{deluxetable}



\end{document}